\documentclass[12pt,tighten,nofootinbib,amssymb,floatfix]{article}

\usepackage{amssymb}
\usepackage{typearea}
\usepackage{t1enc}
\usepackage{mathrsfs}
\usepackage{graphicx}
\usepackage{multirow}
\usepackage{hyperref}
\usepackage{multirow}
\usepackage{amsmath}
\usepackage{arydshln}
\usepackage{tikz}
\usepackage[hang,small,bf]{caption}
\usepackage{ulem}
\usepackage[footskip=0.3in, margin=.85in]{geometry}
\newcommand{\MSb}{{\overline{\rm MS}}}
\renewcommand{\Re}{{\rm Re}\,}
\renewcommand{\Im}{{\rm Im}\,}
\newcommand{\be}{\begin{equation}}
\newcommand{\ee}{\end{equation}}
\newcommand{\bea}{\begin{eqnarray}}
\newcommand{\eea}{\end{eqnarray}}

\begin{document}

\parskip=8pt
\enlargethispage{1\baselineskip}

\title{A complete non-perturbative renormalization prescription for quasi-PDFs}

\author{ \normalsize Constantia Alexandrou$^{a,b}$,
{Krzysztof Cichy$^{c,d}$},
{Martha Constantinou$^e$\footnote{Corresponding author: marthac@temple.edu}},\\
\normalsize
{Kyriakos Hadjiyiannakou$^b$},
{Karl Jansen$^f$},
{Haralambos Panagopoulos$^a$},
{Fernanda Steffens$^f$}}

\date{}
\maketitle

\parskip=8pt

\vspace*{-1cm}
\noindent
\begin{center}
$^a$ {\small\it{Department of  Physics,  University  of Cyprus,  POB  20537,  1678  Nicosia,  Cyprus}}\\
$^b$ {\small\it{The Cyprus Institute, 20 Kavafi Str., Nicosia 2121, Cyprus}} \\
$^c$ {\small\it{Goethe-Universitat Frankfurt am Main, Institut f\"ur Theoretische
Physik, Max-von-Laue-Strasse 1, 60438 Frankfurt am Main, Germany}}\\
$^d$ {\small\it{Faculty of Physics, Adam Mickiewicz University, Umultowska 85, 61-614 Pozna\'{n}, Poland}}\\
$^e$ {\small\it{Department of Physics,  Temple University,  Philadelphia,  PA 19122 - 1801,  USA}}\\
$^f$ {\small\it{John von Neumann Institute for Computing (NIC), DESY, Platanenallee 6, 15738 Zeuthen, Germany}}
\end{center}

\begin{center}
\vspace*{1mm}
\includegraphics[width=0.13\textwidth,angle=0]{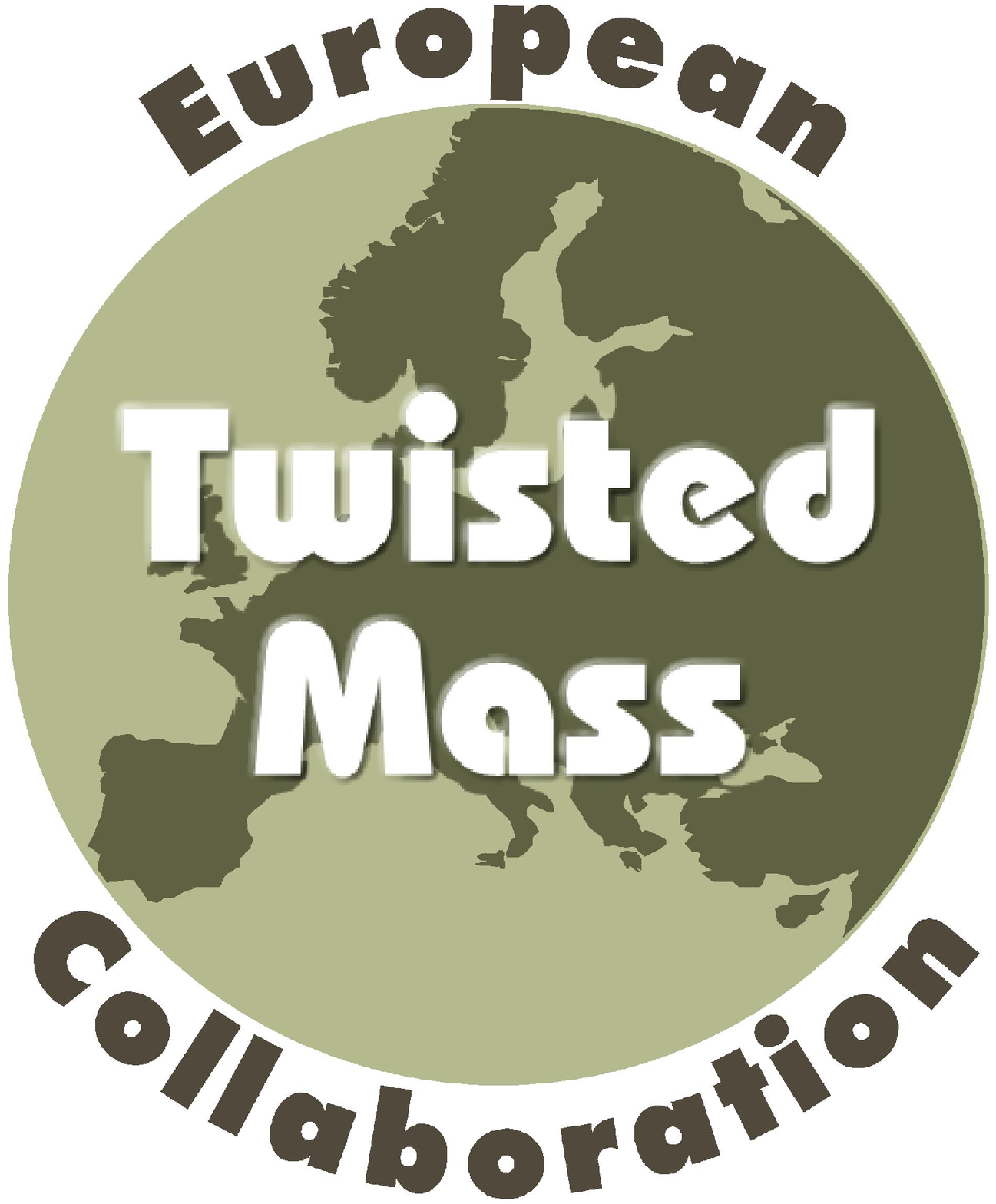}
\end{center}

\vspace*{.75cm}
\abstract{
In this work we present, for the first time, the non-perturbative renormalization for the unpolarized, 
helicity and transversity quasi-PDFs, in an RI$'$ scheme. The proposed prescription addresses 
simultaneously all aspects of renormalization: logarithmic divergences, finite renormalization as well as
the linear divergence which is present in the matrix elements of fermion operators with Wilson lines. 
Furthermore, for the case of the unpolarized quasi-PDF, we describe how to eliminate the unwanted 
mixing with the twist-3 scalar operator.

We utilize perturbation theory for the one-loop conversion factor that brings the renormalization functions
to the $\MSb$-scheme at a scale of 2 GeV. We also explain how to improve the estimates on the 
renormalization functions by eliminating lattice artifacts. The latter can be computed in one-loop perturbation 
theory and to all orders in the lattice spacing. 

We apply the methodology for the renormalization to an ensemble of twisted mass fermions with $N_f{=}2{+}1{+}1$
dynamical quarks, and a pion mass of around 375 MeV. 
}
\vskip .75cm





\newpage
\section{Introduction}
\label{sec:intro}

Parton distribution functions (PDFs) describe the inner dynamics of partons inside a hadron \cite{Feynman:102074}.
They have a non-pertubative nature and, thus, they can not be computed in perturbation theory. Lattice QCD is an
ideal formulation to study the PDFs from first principles, in large scale simulations. 
However, PDFs are usually defined on the light cone, which poses a problem for the standard Euclidean formulation.
Hence,  hadron structure calculations in lattice QCD are related to other quantities  that are accessible in a Euclidean 
spacetime. This led to a long history of investigations of Mellin moments of PDFs and nucleon form factors (see
\cite{Constantinou:2014tga, Constantinou:2015agp,Alexandrou:2015yqa, Alexandrou:2015xts, Syritsyn:2014saa}
for recent reviews). In practice, there are severe limitations in the reconstruction of the PDFs mainly due to the small signal-to-noise ratio for the 
high moments. In addition,  there are inevitable problems with power divergent mixings with lower dimensional operators.
Therefore, the task of reconstructing the PDFs from their moments is practically unfeasible.

\medskip
{\textit{Ab initio}} evaluations of PDFs are of high importance as they would be a stringent test of non-perturbative 
aspects of QCD. The fact that a calculation of the PDFs from first principles is missing is a pressing problem that 
prevents a deeper understanding of the nucleon structure. 
Our current knowledge on the PDFs relies on phenomenological fits from experimental data using perturbation theory.
These parameterized PDFs serve, for example, as input for the computation of cross sections used for
presently running colliders, most notably the LHC, and also to plan future collider experiments which are themselves 
tests of the Standard Model. The parameterizations are, however, not without ambiguities~\cite{Jimenez-Delgado:2013sma}.
In addition, there are kinematical regions that are not experimentally accessible.
The large Bjorken $x$ region is one of them, with large uncertainties most dramatically seen in the down quark 
distributions \cite{Accardi:2016qay,Alekhin:2017kpj}. The transversity PDF is yet another example of a PDF that is 
only poorly constrained by phenomenology.

\medskip
A pioneering method for a direct computation has been suggested by X.~Ji \cite{Ji:2013dva}.
In this approach, instead of matrix elements defined on the light cone, one calculates matrix elements 
of fermion operators including a finite-length Wilson line, and whose Fourier transform defines the so-called 
quasi-PDFs. This is achieved by taking the Wilson line in a purely spatial direction, conventionally chosen 
to be the $z$-direction, instead of the $+$-direction on the light cone.
In terms of hadron kinematics, the nucleon momentum is usually taken along this spatial direction. 
At large, but finite momenta, the quasi-PDFs can then be related to the light-front 
PDFs via a matching procedure~\cite{Xiong:2013bka,Chen:2016fxx}. Ji's approach has already been tested in 
Refs.~\cite{Lin:2014zya,Gamberg:2014zwa,Alexandrou:2015rja,Chen:2016utp,Alexandrou:2016jqi} and all these results 
are promising, i.e.\ they give a correct shape of the PDFs after the matching procedure. Certain properties of quasi-PDFs, 
like the nucleon mass dependence and target mass effects, have also been analyzed via their relation with transverse momentum 
dependent distribution functions (TMDs)~\cite{Radyushkin:2017ffo,Radyushkin:2016hsy}. 
Note also the appearance of a related approach, pseudo-PDFs, which is a different generalization of light-cone PDFs to finite nucleon momenta \cite{Radyushkin:2017cyf,Orginos:2017kos}.
Refs.~\cite{Carlson:2017gpk,Briceno:2017cpo} discussed the role of the Euclidean signature in the computation of quasi-PDFs, as compared to light-front PDFs, and the latter reference proved that matrix elements obtained from Euclidean lattice QCD are identical to those obtained using the LSZ reduction formula in Minkowski space. 
Nevertheless, the matrix elements of quasi-PDFs contain divergences that need to be eliminated via renormalization in order to obtain 
meaningful results that can be compared to the physical PDFs.

\medskip
To date, all works on the quasi-PDFs only considered the bare matrix elements, as the renormalization process is
highly non-trivial and was not addressed until recently. In particular, new complications arise in the 
renormalization of the Wilson line operators, compared to the local operators. For one, in addition to the 
logarithmic divergences there is a linear divergence~\cite{Dotsenko:1979wR} with respect to the lattice regulator, 
$a$, that prohibits one to take the continuum limit prior to its elimination. To one-loop level in perturbation theory the 
divergence is manifesting itself as a linear divergence, computed in Ref.~\cite{Constantinou:2017sej} for a variety 
of fermion and gluon actions. However, it is of utmost importance to extract the power divergence non-perturbatively, 
which is one of the goals of this work.  Another feature of these operators that brings in new complications is the fact 
that certain choices of the Dirac structure exhibit mixing~\cite{Constantinou:2017sej}.

\medskip
To show that these matrix elements can be renormalized, in particular that the linear divergence associated with
the Wilson line can be eliminated, is of paramount importance. Without renormalization, the whole quasi-PDF strategy
is incomplete and unable to provide any useful information to the theoretical and experimental community. Some suggestions
for the elimination of the linear divergence via the static potential were proposed in Refs.~\cite{Ma:2014jla,Ishikawa:2016znu}.
In Ref.~\cite{Chen:2016fxx} a one loop calculation of the linear divergence has been made, and that motivated the definition of an
improved quasi-PDF that is free of power divergences. One has, nevertheless, to show that such a procedure can be done non-perturbatively.
Another method to extract the coefficient of the linear divergence using the nucleon matrix elements of the quasi-PDFs was also presented 
in Ref.~\cite{Constantinou:2017sej}. An alternative technique to suppress the linear divergence was discussed in Ref.~\cite{Monahan:2016bvm} 
utilizing the gradient flow. Very recently, two papers discussed the employment of the auxiliary field formalism for the renormalization of 
quasi-PDFs~\cite{Ji:2017oey,Green:2017xeu}, where the Wilson line is replaced by a Green's function of the introduced auxiliary field. 
The renormalizability of quasi-PDFs to all orders in perturbation theory was addressed in Ref.~\cite{Ishikawa:2017faj}.

\medskip
In this paper we propose, for the first time\footnote{After the submission of our work, the proposed renormalization programme was also applied in Ref.~\cite{Chen:2017mzz}.}, a concrete renormalization method of the quasi-PDFs in a fully non-perturbative manner.
We provide the prescription of the method and show examples of the renormalized matrix elements. We also discuss the
elimination of the mixing between the unpolarized quasi-PDF and the twist-3 scalar operator.
We employ the RI$'$ renormalization scheme~\cite{Martinelli:1994ty} and we convert the results to the
$\MSb$ scheme at a reference scale, $\bar\mu{=}2$ GeV, using the one-loop conversion factor computed in Ref.~\cite{Constantinou:2017sej}.
As a test case, we focus on the helicity quasi-PDF to demonstrate results, as it is free of mixing.

\medskip
The outline of the paper is as follows: In Section 2 we provide the theoretical setup related to the nucleon
matrix elements and the renormalization prescription in the presence and absence of mixing,
as well as the data on the conversion factor computed perturbatively.
Section 3 includes results on the
renormalization functions, a discussion on the systematic uncertainties in the $Z$-factors, renormalized matrix elements for the helicity case, as well as results of the matching to light-front PDFs. Finally we conclude and give our future directions.

\section{Theoretical setup}

In this section we briefly introduce the nucleon matrix elements for the quasi-PDFs that we aim to renormalize. 
We also explain the renormalization prescription for the three types of PDFs: unpolarized, helicity and transversity.
For details on the computation of the nucleon matrix elements, we refer to Refs.~\cite{Alexandrou:2015rja,Alexandrou:2016jqi}.

\vspace*{0.25cm}
\subsection{Nucleon matrix elements}
\vspace*{0.15cm}

We consider matrix elements of non-local fermion operators that contain a straight Wilson line, denoted by
$h_\Gamma(P_3,z)$. The variable $z$ is the length of the Wilson line, and $P_3$ is the nucleon momentum, 
which is taken in the same direction as the Wilson line.  
The quasi-PDFs can be computed from the Fourier transform of the following local matrix elements:
\begin{equation}
\label{MElements}
h_\Gamma(P_3,z)=\langle N |\bar{\psi}(0,z)\,\Gamma \,W_3(z) \psi(0,0) |N\rangle,
\end{equation}
where $|N\rangle$ is a nucleon state with spatial momentum $\vec{P}{=}(0,0,P_3)$ along the $3$-direction and 
$W_3(z)$ is a Wilson line of length $z$ in the same direction. $\Gamma$ denotes the Dirac structure of the
operator insertion, which is $\gamma_\mu$ (unpolarized), $\gamma_\mu\cdot\gamma_5$ (helicity), $\sigma_{\mu \nu}$ 
(transversity). In the works appearing in the literature $\gamma_\mu$ is taken along the Wilson line. In principle, one may 
choose $\gamma_\mu$ orthogonal to the direction of the Wilson line.  In this case, the unpolarized operator is 
free of mixing, while the helicity and transverity do mix. For example, choosing the $\gamma$-matrix in the temporal direction 
is important for a faster convergence to the physical PDFs, as discussed in Ref.~\cite{Radyushkin:2016hsy}. Our recent 
work~\cite{Constantinou:2017sej} indicates that $\vec P_3 {\perp} \vec z$ \,\,($\vec P_3 {\parallel} \vec z$) is ideal for 
the unpolarized (helicity and transversity) case.

\medskip
To calculate the bare matrix elements, we use the setup of Ref.~\cite{Alexandrou:2016jqi}.
We consider one ensemble of dynamical $N_f{=}2{+}1{+}1$ twisted mass fermions produced by ETMC~\cite{Baron:2010bv}, with volume 
$32^3 {\times} 64$, lattice spacing $a {\approx}0.082$\,fm~\cite{Carrasco:2014cwa} and a bare twisted mass of $a\mu {=} 0.0055$, 
which corresponds to a pion mass of around 375 MeV. We performed our calculations on 1000 gauge configurations with 15 
forward propagators and 2 stochastic propagators, i.e.\ 30000 measurements in total. We will present results for momentum $P_3=\frac{6\pi}{L}$,
which is around 1.4 GeV in physical units. Gaussian smearing has been employed on the nucleon interpolating fields in the calculation of the matrix elements
\cite{Alexandrou:2016jqi}.

\vspace*{0.25cm}
\subsection{Renormalization scheme}
\label{sec:scheme}
\vspace*{0.15cm}

Here we discuss a fully non-perturbative renormalization prescription that will remove all divergences inherited in the matrix elements 
of the quasi-PDFs, as well as the mixing, as indicated in the perturbative analysis of Ref.~\cite{Constantinou:2017sej}. In a nutshell, the proposed renormalization program: 
\begin{itemize}
\item[\bf 1] removes the linear divergence that resums into a multiplicative exponential factor, 
$e^{-\delta m |z|/a + c |z|}$. The coefficient $\delta m$ represents the strength of the divergence and is 
expected to be operator independent, as it is related only to the Wilson line. $c$ is an arbitrary scale
\cite{Sommer:2015hea} that can be fixed by such a renormalization prescription;
\item[\bf 2] takes away the logarithmic divergence with respect to the regulator, $\log(a\bar\mu_0)$, where $\bar\mu_0$ is 
the RI$'$ renormalization scale;
\item[\bf 3] applies the necessary finite renormalization related to the lattice regularization;
\item[\bf 4] eliminates the mixing that appears in the unpolarized operator, as the bare matrix element is
a linear combination of the unpolarized quasi-PDF and the twist-3 scalar operator. The two may be disentangled by
the construction of a $2{\times}2$ mixing matrix.
\end{itemize}

\medskip
We adopt a renormalization scheme which is applicable non-perturbatively, that is, the RI$'$~scheme \cite{Martinelli:1994ty}. 
We compute vertex functions of the operators under study, between external quark states, with the setup being in momentum 
space, and the operator defined as:
\be
{\cal O}_\Gamma = \overline \psi(x) \, \Gamma \,\mathcal{P}\, e^{i\,g\,\int_{0}^{z} A(\zeta) d\zeta}\, \psi(x+z\hat{\mu})\,,
\ee
where $\Gamma=\gamma_\mu,\,\gamma_\mu\cdot\gamma_5,\,\sigma_{\mu\nu}$ ($\nu\neq\mu$). The path ordering of the exponential 
appearing in the above expression becomes, on the lattice, a series of path ordered gauge links. The renormalization functions
($Z$-factors) depend on the length of the Wilson line and, thus, we perform a separate calculation for each value of $z$. Typically, $z$
goes up to half of the spatial extent of the lattice.

\medskip
The renormalization prescription is along the lines of the program developed for local operators and the construction of the vertex functions is
described in Ref.~\cite{Alexandrou:2010me}. The difference between the renormalization of the local operators and
the Wilson-line operators is the linear divergence that appears in the latter case. However, there is no need to separate this divergence
from the multiplicative renormalization and, therefore, the technique described below may successfully extract both contributions
at once. 

\vspace*{0.55cm}
\centerline{\bf\underline{\textit{Helicity and transversity quasi-PDFs}}}
\vskip 0.25cm

We first provide the methodology for a general operator with a Wilson line in the absence of any mixing. This is applicable for the 
helicity and transversity quasi-PDFs, provided that their Dirac structure is chosen along the Wilson line. The renormalization functions of the 
Wilson-line operators, $Z_{\cal O}$, 
are extracted by imposing the following conditions:
\be
Z_q^{-1}\,Z_{\cal O}(z)\,\frac{1}{12} {\rm Tr} \left[{\cal V}(p,z) \left({\cal V}^{\rm Born}(p,z)\right)^{-1}\right] \Bigr|_{p^2=\bar\mu_0^2} = 1\, ,
\label{renorm}
\ee
where $Z_q$ is the renormalization function of the quark field obtained via
\be
Z_q = \frac{1}{12} {\rm Tr} \left[(S(p))^{-1}\, S^{\rm Born}(p)\right] \Bigr|_{p^2=\bar\mu_0^2}  \,.
\label{Zq_cond}
\ee
The trace is taken over spin and color indices, and the momentum $p$ entering the vertex function is set to the RI$'$ renormalization scale 
$\bar\mu_0$. In Eq.~(\ref{renorm}) ${\cal V}(p,z)$ is the amputated vertex function of the operator and ${\cal V}^{{\rm Born}}$ is its 
tree-level value, i.e. ${\cal V}^{{\rm Born}}(p,z) {=} i\gamma_3\gamma_5 \,e^{i p z}$ for the helicity operator. Also, $S(p)$ is the fermion 
propagator and $S^{{\rm Born}}(p)$ is its tree-level. The RI$'$ scale $\bar\mu_0$ is chosen such that its $z$-component is
the same as the momentum of the nucleon. Such a choice serves as a suppression of discretization effects, as different classes of spatial 
components have different discretization effects, and scales of the form $(n_t,3,3,3)$ have small discretization effects
\cite{Alexandrou:2015sea}. We test both diagonal (democratic) and parallel momenta to the Wilson line (in the spatial direction), that is 
$a\vec{\bar{\mu}}_0{=} \frac{2\pi}{L}\,(P_3,P_3,P_3)$ and $a\vec{\bar{\mu}}_0{=} \frac{2\pi}{L}\,(0,0,P_3)$, respectively. We will refer 
to these choices as ``diagonal'' and ``parallel''. The latter are expected to have larger lattice artifacts, as the ratio 
\be
\label{Phat}
\hat{P}{\equiv}\frac{\sum_\rho \bar\mu_{0_\rho}^4}{\left(\sum_\rho \bar\mu_{0_\rho}^2\right)^2}
\ee
 is higher than for diagonal momenta. Using renormalization scales leading to a small value for such a ratio has been successful for the local 
fermion operators~\cite{Constantinou:2010gr,Alexandrou:2010me}.

\medskip
The vertex functions ${\cal V}(p)$ contain the same linear divergence as the nucleon matrix elements. 
This is crucial as it allows the extraction of the exponential together with the multiplicative $Z$-factor, 
through the renormalization condition of Eq.~(\ref{renorm}). Regarding the renormalizability of quasi-PDFs, 
it was proven to be multiplicative to all orders in Refs.~\cite{Ishikawa:2016znu,Ishikawa:2017faj}. 
The authors consider quasi-PDFs defined in coordinate space and prove the multiplicative renormalizability, 
and the same holds for the definition in Eq.~(\ref{MElements}). This is due to the fact that the former require, 
in addition, a consistent subtraction scheme to remove all terms divergent in the limit $\xi_z\rightarrow 0$. 
Moreover, the renormalization of bilocal composite operators is also studied in Ref.~\cite{Ishikawa:2016znu,Ishikawa:2017faj}. 
Based on the above, $Z_{\cal O}$ can be factorized as
\be
Z_{\cal O}(z) = \overline{\cal Z}_{\cal O} \,\, e^{+\delta m |z|/a - c |z|}\,,
\ee
where $\overline{\cal Z}_{\cal O}$ is the multiplicative $Z$-factor of the operator and $\delta m$ is the strength of the linear divergence.
The exponential includes a term with an arbitrary scale $c$ that could be of the form $c|z|$~\cite{Sommer:2015hea}. Note that the exponential comes with a 
different sign compared to the nucleon matrix element, as $Z_{\cal O}$ is related to the inverse 
of the vertex function,
\be
Z_{\cal O} = \frac{Z_q}{\frac{1}{12} {\rm Tr} \left[{\cal V}(p) \left({\cal V}^{\rm Born}(p)\right)^{-1}\right] \Bigr|_{p=\bar\mu} }\,.
\ee
Such a construction of the $Z$-factor justifies the reason why the elimination of the power divergence in the nucleon matrix elements is successful
by multiplying with $Z_{\cal O}$, provided that it has been calculated on the same ensemble.
Consequently, the above renormalization condition handles all the divergences which are present in the matrix element under 
consideration. Note that in the absence of a Wilson line ($z{=}0$), the $Z$-factors reduce to the ones for local currents,
which are free of any power divergence. For example, for the helicity operator, $Z_A(z{=}0)$ is the standard $Z$-factor 
of the axial current.

\medskip
We would like to point out that knowledge of the coefficient $\delta m$ provides insight on the strength of the power divergence.
One can pursue this direction via the static potential~\cite{Ishikawa:2016znu} or the technique proposed in Ref.~\cite{Constantinou:2017sej}. 
If the linear divergence is extracted, one may apply the matching of Ref.~\cite{Chen:2016fxx}, which includes the coefficient $\delta m$.
However, there is still a necessity to compute the multiplicative $Z$-factor to cure any logarithmic divergences, apply finite 
renormalization, as well as fix the arbitrary scale $c$.

\vspace*{0.75cm}
\centerline{\bf\underline{\textit{Unpolarized quasi-PDF}}}
\vskip 0.25cm

The case of the unpolarized quasi-PDF requires special treatment, if the Dirac structure is in the same direction
as the Wilson line. As demonstrated in Ref.~\cite{Constantinou:2017sej}, for such a choice there is a mixing with 
the twist-3 scalar operator\footnote{For twisted mass fermions the mixing is between the vector and pseudoscalar 
currents.}, that must be eliminated. This mixing appears in some lattice regularizations (in particular, Wilson and twisted mass fermions) due to the breaking of chiral symmetry, and is found to be finite. 
We establish notation by using the subscripts $S$ and $V$ for the scalar and vector (unpolarized) operators, respectively. 
The corresponding operators are:
\bea
{\cal O}_S \hspace*{-0.15cm}&=&\hspace*{-0.15cm} \overline \psi(x) \, \hat{1} \,\mathcal{P}\, e^{i\,g\,\int_{0}^{z} A(\zeta) d\zeta}\, \psi(x+z\hat{\mu})\,, \\[2ex]
{\cal O}_V \hspace*{-0.15cm}&=&\hspace*{-0.15cm} \overline \psi(x) \, \gamma_\mu \,\mathcal{P}\, e^{i\,g\,\int_{0}^{z} A(\zeta) d\zeta}\, \psi(x+z\hat{\mu})\,,
\eea
and we represent their nucleon matrix elements as $h_S(P_3,z)$ and $h_V(P_3,z)$.
To disentangle the two contributions from their bare matrix elements, one must compute the multiplicative renormalization 
and mixing coefficients from the following $2{\times}2$ matrix:
\begin{equation}
  \binom{{\cal O}^R_V(P_3,z)}{{\cal O}^R_S(P_3,z)} = \hat{Z}(z)\cdot
  \binom{{\cal O}_V(P_3,z)}{{\cal O}_S(P_3,z)}\,,
\end{equation}
where $\hat{Z}$ is the matrix of the mixing between the scalar and vector operators, that is
\begin{equation}
 \hat{Z}(z) = \begin{pmatrix} Z_{VV}(z) & Z_{VS}(z) \\ Z_{SV}(z) & Z_{SS}(z) \end{pmatrix} \,.
\end{equation}
According to the above mixing, the renormalized unpolarized quasi-PDF, $h^R_V(P_3,z)$, is related to the bare scalar and unpolarized via:
\begin{equation}
h^R_V(P_3,z) = Z_{VV}(z) \,\, h_V(P_3,z) + Z_{VS}(z) \,\, h_S(P_3,z) \,,
\label{h_R}
\end{equation}
where $Z_{VV}$ and $Z_{VS}$ are computed in a certain scheme. In the present work
we employ the RI$'$ scheme and then convert to the $\MSb$ scheme, at an energy scale $\bar\mu$.
The $Z_{ii}$ factors can be computed following a prescription similar to Eq.~(\ref{renorm}), that is:
\be
\label{renormMIX}
Z_q^{-1}\,\hat{Z}(z)\, {\cal \hat{V}}(p,z)\Bigr|_{p=\bar\mu} = \hat{1}\,,
\ee
where the elements of the vertex function matrix ${\cal \hat{V}}$ are given by the trace
\be
\left({\cal \hat{V}}(z) \right)_{ij}=  \frac{1}{12} {\rm Tr} \left[{\cal V}_i(p,z) \left({\cal V}_j^{\rm Born}(p,z)\right)^{-1}\right]\,,\quad i,j=S,V\,.
\ee
In the above equation ${\cal V}_i^{\rm Born}$ is the tree-level expression of the operator ${\cal O}_i$.
Thus, all matrix elements of $\hat{Z}$ can be extracted by a set of linear equations, which can be written in 
the following matrix form:
\be
Z_q^{-1}\,\begin{pmatrix} Z_{VV}(z) & Z_{VS}(z) \\[2ex] Z_{SV}(z) & Z_{SS}(z) \end{pmatrix}  \cdot
\begin{pmatrix}  \left({\cal \hat{V}}(z) \right)_{VV} & \left({\cal \hat{V}}(z) \right)_{SV} \\[2ex]
 \left({\cal \hat{V}}(z) \right)_{VS} &\left({\cal \hat{V}}(z) \right)_{SS} \end{pmatrix} =
\begin{pmatrix} 1& 0 \\[2ex] 0 & 1\end{pmatrix} \,.
\ee
As can be seen from Eq.~(\ref{h_R}), a major component in the renormalization of 
the unpolarized quasi-PDF is knowledge of the scalar nucleon matrix element $h_S(P_3,z)$. To date, no lattice calculation
is available for $h_S(P_3,z)$, as a mixing was not anticipated prior the work of Ref.~\cite{Constantinou:2017sej}. Therefore,
a proper renormalization of the unpolarized quasi-PDF is still pending. However, the mixing appears to be greatly suppressed
in the presence of a clover term in the fermionic action. This is of high importance, as we are currently computing the quasi-PDFs on
an ensemble of twisted mass clover improved fermions at the physical pion mass~\cite{Abdel-Rehim:2015pwa,Abdel-Rehim:2015owa}.

\vspace*{0.25cm}
\subsection{Conversion to the $\MSb$-scheme}
\vspace*{0.15cm}

The fermion operators with a finite-length Wilson line are scheme and scale dependent. As a result, having obtained 
the $Z$-factors in the RI$'$ scheme as depicted in Eqs.~(\ref{renorm}) and (\ref{renormMIX})
at the RI$'$ scale $\bar\mu_0$, we must convert them to the $\MSb$ scheme at a scale $\bar\mu$. This conversion factor has 
been computed in one-loop continuum perturbation theory in Ref.~\cite{Constantinou:2017sej}. In addition, comparison 
with phenomenological estimates is done typically at $\bar\mu{=}2$ GeV. This defines the value of the $\MSb$ renormalization 
scale chosen in the expressions for the conversion factors. Alternatively, one can choose $\bar\mu{=}\bar\mu_0$, and then
evolve the scale to 2 GeV, which requires the anomalous dimension of the operator:
\be
\gamma_{\cal O} =  -3 \frac{g^2\,C_f}{16\pi^2} \,.
\ee
Note that to one-loop level, the anomalous dimension does not dependent on the Dirac structure of the operator
and is the same in the RI$'$ and $\MSb$ schemes. The evolution to the $\MSb$ renormalization scale  
$\bar\mu{=}2$ GeV can be performed using the intermediate Renormalization Group Invariant scheme (RGI), as 
employed in Refs.~\cite{Constantinou:2014fka,Alexandrou:2015sea}. In the framework of this paper we tested both 
methods and their difference was found to be negligible.

\begin{figure}[h]
\centering
\includegraphics[scale=0.265,angle=-90]{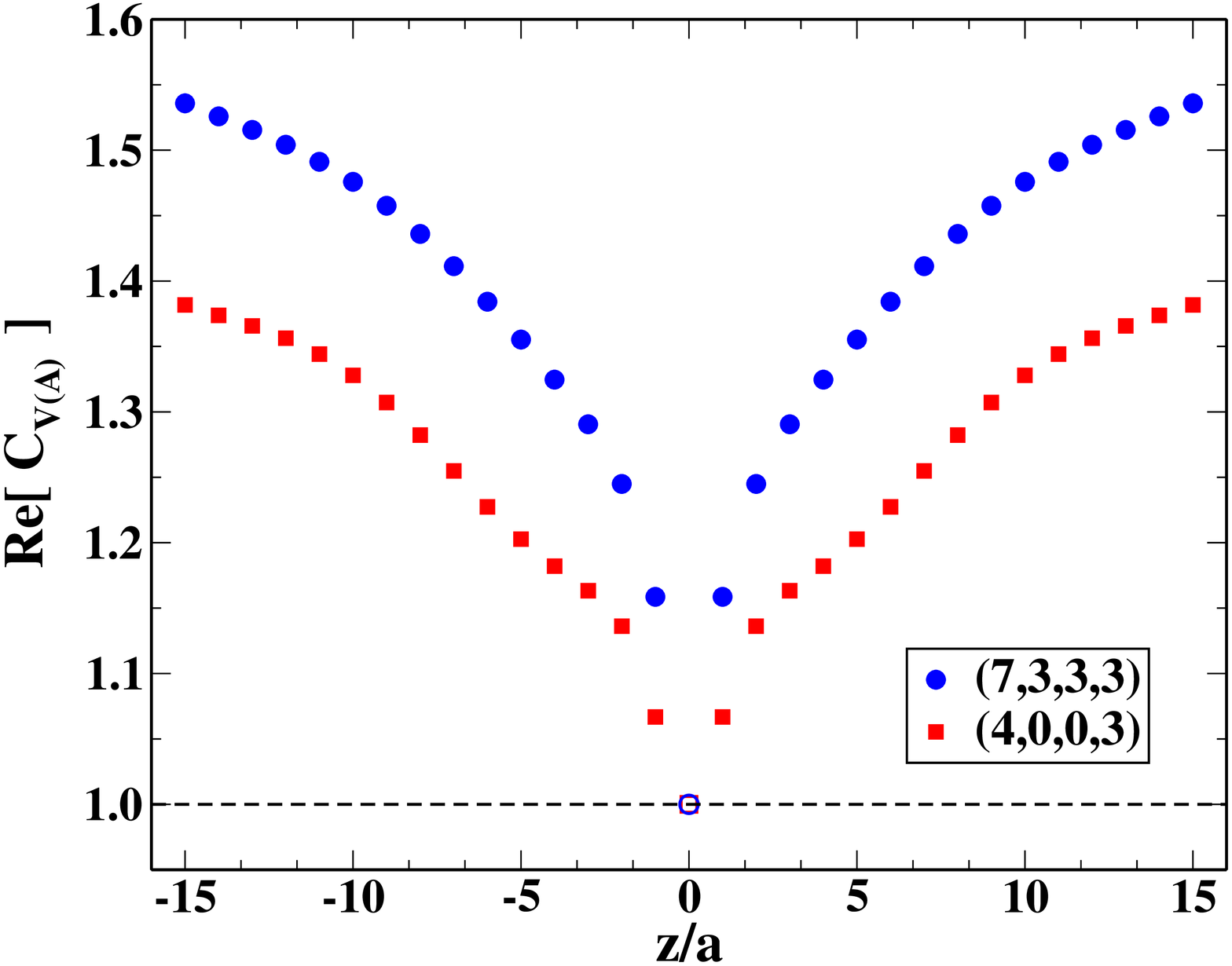}\,\,\,
\includegraphics[scale=0.265,angle=-90]{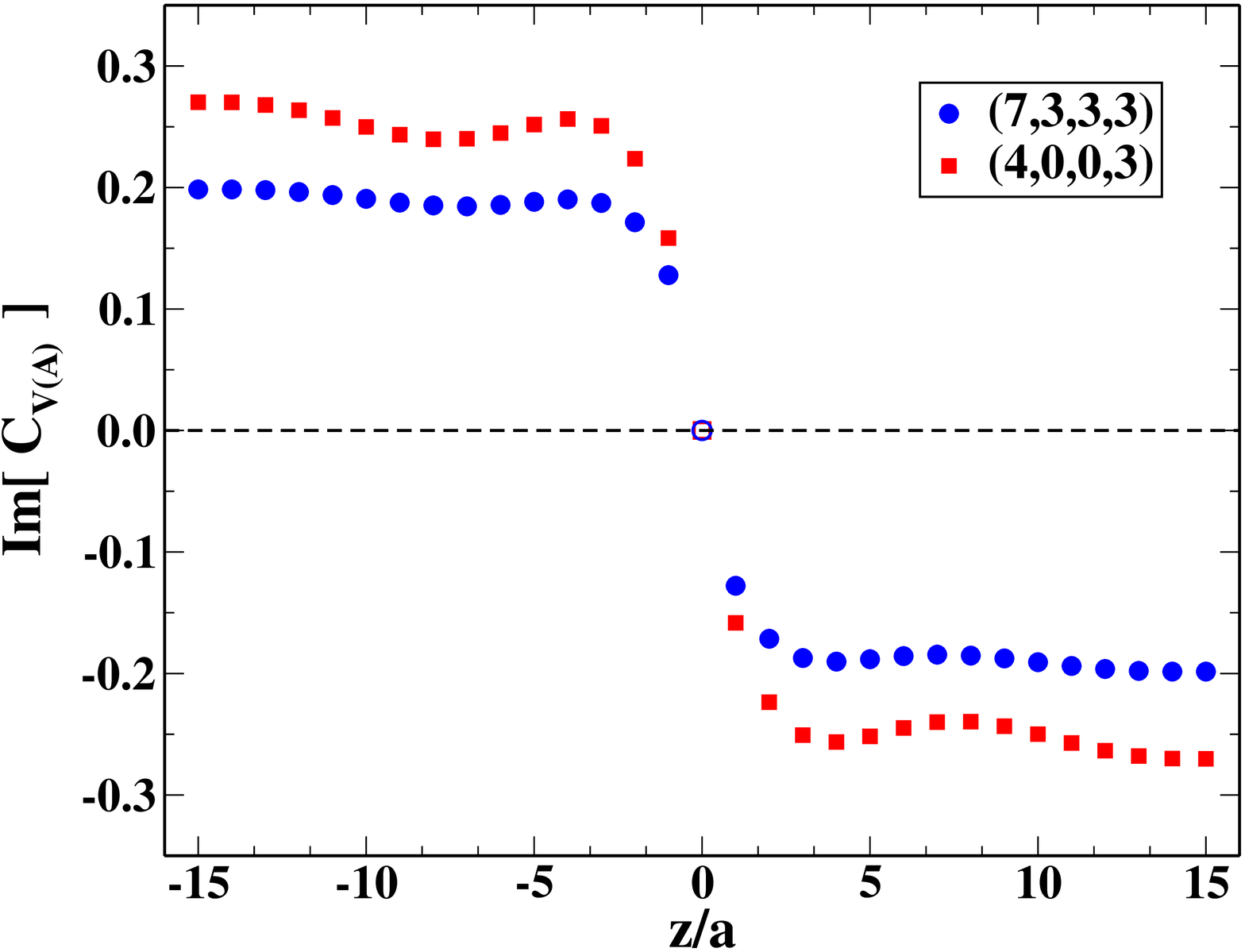}
\vspace*{-0.3cm}
\begin{minipage}{15cm}
\hspace*{3cm}
\caption{\small{One-loop conversion factor from the RI$'$ to the $\MSb$ scheme for $\bar\mu{=}\bar\mu_0$. In 
the left (right) plot we show the real (imaginary) part of the conversion factor as a function 
of the length of the Wilson line in lattice units. Two choices of the RI$'$ scale have been
employed: $a\bar\mu_0{=}\frac{2\pi}{32}\, (\frac{7}{2}{+}\frac{1}{4}, 3,3,3)$ (blue circles)  and 
$a\bar\mu_0{=}\frac{2\pi}{32}\, (\frac{4}{2}{+}\frac{1}{4}, 0,0,3)$  (red squares). 
We use the abbreviation (7,3,3,3) and (4,0,0,3) in the legends, respectively. }}
\label{fig:C} 
\end{minipage}
\end{figure}

\medskip
As discussed in the previous section, we employ two types of the RI$'$ renormalization scale regarding the spatial direction: 
one the same as in the nucleon matrix elements, $a\vec{\bar{\mu}}_0{=}\frac{2\pi}{L}\,(0,0,P_3)$, (parallel to the Wilson line direction), 
and one which is diagonal and each direction has a value of $P_3$, that is $a\vec{\bar{\mu}}_0{=}\frac{2\pi}{L}\,(P_3,P_3,P_3)$. 
The conversion factor depends on both the RI$'$ and $\MSb$ renormalization scales. While $\bar\mu$ is typically fixed to 2 GeV, 
$\bar\mu_0$ can change affecting the numerical values of the conversion factor. Such a case is illustrated in Fig.~\ref{fig:C} 
for the unpolarized and helicity operators, which share the same conversion factor\footnote{The one-loop calculation does not 
depend on the prescription which one adopts for extending $\gamma_5$ to $D$ dimensions (see, e.g.,
Refs.~\cite{Buras:1989xd,Patel:1992vu,Larin:1993tp,Larin:1993tq,Skouroupathis:2008mf,Constantinou:2013pba} 
for a discussion of four relevant prescriptions and some conversion factors among them).}. 
We focus on the renormalization of the matrix elements with the nucleon boosted by $a P_3{=}\frac{6\pi}{L}$, and we apply the same
to the conversion factor, for the two cases of the renormalization scale. For the specific value of $P_3$, we choose the temporal
direction of $\bar\mu_0$ such that the ratio $\hat{P}$ defined above, is as small as possible in order to suppress
lattice artifacts. Nevertheless, for the ``parallel'' case the minimum value for the ratio is $\hat{P}{=}0.54$ which is already very high.
The chosen values for $a\bar\mu_0$ are: $\frac{2\pi}{32}\, (\frac{7}{2}{+}\frac{1}{4}, 3,3,3)$ and 
$\frac{2\pi}{32}\, (\frac{4}{2}{+}\frac{1}{4}, 0,0,3)$ for the ``diagonal'' and ``parallel'' case, respectively. 

\medskip
The real and imaginary parts of the conversion factor are plotted as a function of the length of the Wilson 
line, $z$. We stress that the dependence of the conversion factor on the renormalization scales and the length of the 
Wilson line is highly non-trivial and is expressed in terms of integrals of modified Bessel functions. Consequently, the data points
shown in Fig.~\ref{fig:C} have been computed numerically for the specific scales, at each value of $z$ separately. We observe
that the real part of the conversion is an order of magnitude larger than the imaginary part. The real part consistently increases 
for increasing values of $z$, while the imaginary part almost immediately stabilizes when $z$ becomes non-zero. Also, the conversion 
factor at $z{=}0$ is equal to unity, as it corresponds to the local vector and axial currents. Note that the case $z{=}0$ is not extracted
from the calculation of Ref.~\cite{Constantinou:2017sej} as it is strictly for $z{\neq}0$\,: the appearance of contract terms beyond
tree level renders the limit $z\to 0$ nonanalytic. On the contrary, the non-perturbative prescription of the previous section
can be applied for any value of $z$, as the calculation is performed on each $z$ separately. The values of Fig.~\ref{fig:C} will be used
in the following section to bring the RI$'$ non-perturbative $Z$-factors to the $\MSb$ scheme. To reliably extract 
the $Z$-factors we have extend the calculation including several values of $\bar\mu_0$ as explained in Section~\ref{Sec3}. The conversion
factor was found to have the same qualitative behavior for all values of $\bar\mu_0$.

\section{Results}
\label{Sec3}

In this section we apply the prescription suggested above and we present our results for the non-perturbative $Z$-factors 
both in the RI$'$ and the $\MSb$ schemes. For demonstration purposes we focus on the data with 5 steps of Hypercubic (HYP) smearing that suppress 
the power divergence and bring the results closer to renormalized nucleon matrix elements. We focus on the multiplicative renormalization 
for the helicity quasi-PDF, and only briefly discuss the case of the unpolarized operator. 

\subsection{RI$'$ scheme and conversion to the $\MSb$ scheme}

As a starting point, we have applied the two values of the RI$'$ scale $\bar\mu_0$ 
used in Fig.~\ref{fig:C} (``parallel'' and ``diagonal''). After converting both cases to the $\MSb$ scheme at $\bar\mu{=}2$ GeV,
we can quantify the systematic uncertainties related to lattice artifacts and truncation of the conversion factor, as explained in the next subsection. 

\begin{figure}[h]
\centering
\includegraphics[scale=0.275,angle=-90]{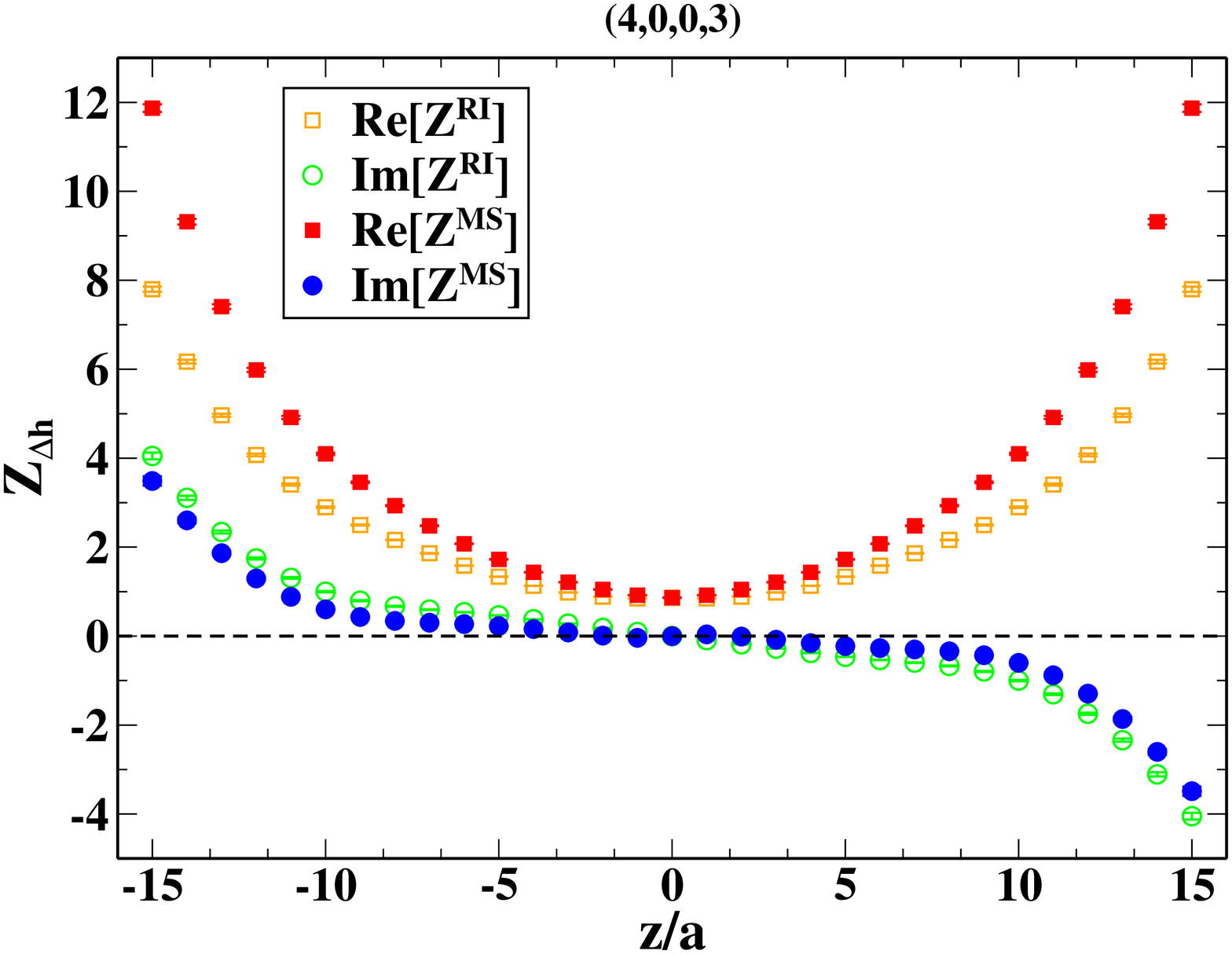}\,\,\,
\includegraphics[scale=0.275,angle=-90]{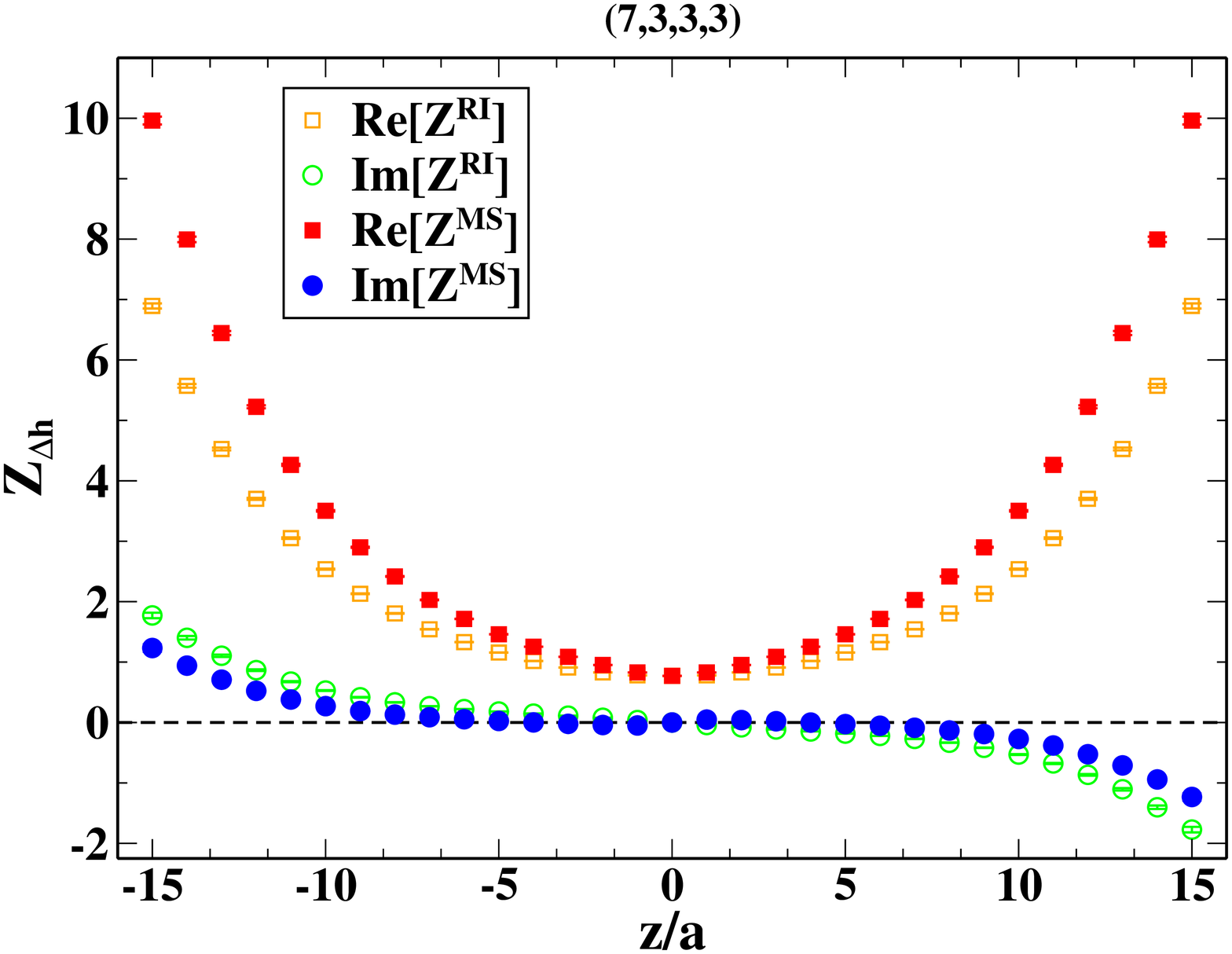}
\vspace*{-0.5cm}
\begin{minipage}{15cm}
\hspace*{3cm}
\caption{\small{The $z$-dependent renormalization function for the matrix element $\Delta h(P_3,z)$ 
with $a P_3=\frac{6\pi}{L}$. The ``parallel'' and ``diagonal'' choices for $\vec{\bar{\mu}}_0$ are shown in the left and right
plots, respectively. Open (filled) symbols correspond to the RI$'$ ($\MSb$) estimates.}}
\label{fig:ZA}
\end{minipage}
\end{figure}

In Fig.~\ref{fig:ZA}, we show the extracted values for the helicity $Z$-factor, $Z_{\Delta h}$, that renormalizes the bare matrix element 
$\Delta h(P_3,z)$. In each plot we overlay the results for the RI$'$ (open symbols) and the $\MSb$ (filled symbols) schemes,
for the real and imaginary part of the $Z$-factor. We employ the momentum source technique~\cite{Gockeler:1998ye,Alexandrou:2015sea} that 
offers high statistical accuracy with a small number of measurements, typically of ${\cal O}(10)$. As can be seen in the plots, the statistical
uncertainties are almost invisible. The left (right) plot corresponds to the ``parallel'' (``diagonal'') choices for $\bar\mu_0$.  
We find that the imaginary part of $Z_{\Delta h}^{\overline{\rm MS}}$ is reduced compared to its counterpart in $Z_{\Delta h}^{\rm RI'}$, and is also rather small 
for low values of $z$. This is more pronounced in the right plot of Fig.~\ref{fig:ZA} for ``diagonal'' $\bar\mu_0$, for which $\Im[Z_{\Delta h}^{\MSb}]$ 
(blue circles) is smaller than the corresponding data from the ``parallel'' scale, especially for large values of $z$.
It is worth mentioning that the perturbative $Z$-factor in dimensional regularization and in the $\MSb$ scheme is real to 
all orders in perturbation theory, as it is extracted only from the poles. Therefore, it is expected that the imaginary part of the
non-perturbative estimates should be highly suppressed. The behavior of the ``diagonal'' scale is encouraging,
as the imaginary part is very close to zero for $|z|$ up to ${\sim}10\,a$.
\vspace*{0.25cm}

\vspace*{0.35cm}
As we have done in previous applications of the RI$'$ scheme to extract the renormalization functions of ultra-local operators
(see, e.g., Ref.~\cite{Alexandrou:2015sea}), here we also use a range of values for the RI$'$ renormalization scale, 
$\bar\mu_0$. This allows us to study the scale dependence of the vertex functions with the external
momentum and identify the window in  which renormalization factors can be extracted as reliably as possible.
Unlike the case of the local currents, the computation of the lattice artifacts to ${\cal O}(g^2\,a^\infty)$
is extremely laborious and not available yet. Thus, one has to be careful with the choices for $\bar\mu_0$, 
as we want to avoid non-perturbative contaminations, and satisfy the criterion that $\hat{P}$ (Eq.~(\ref{Phat})) is small, 
preferably below 0.3, to avoid enhanced cut-off effects. 
We compute the renormalization functions using the values reported in Table~\ref{tab1}, from which we choose an optimal 
range for the fit using the diagonal choices. The parallel $\bar\mu_0$ have only been used to explore the 
systematic uncertainties in Subsection~\ref{sub3.2}. 

{\small{
\begin{table}[!h]
\vspace*{-0.25cm}
\begin{center}
\begin{tabular}{ccccc}
\hline
\hline\\[-2ex]
Label & $(n_t,n_x,n_y,n_z)$   &  $(a \bar\mu_0)^2$ &  $\bar\mu$ (GeV)  &  $\hat{P}$  \\[0.5ex]
\hline
\hline\\[-2ex]
& & diagonal & & \\
\hline\\[-2ex]
 m1 &(4,3,3,3)		&1.236       &2.671	 &0.261  		\\
 m2 &(5,3,3,3)		&1.332       &2.773	 &0.251 		\\
 m3 &(6,3,3,3)		&1.448       &2.891	 &0.251		\\
 m4 &(7,3,3,3)		&1.583       &3.023	 &0.261 		\\
 m5 &(8,3,3,3)		&1.737       &3.167	 &0.280		\\
 m6 &(9,3,3,3)		&1.911        &3.321	 &0.306 		\\
 m7 &(10,3,3,3)	&2.104       &3.484 	 &0.339		\\
 m8 &(11,3,3,3)	&2.316       &3.656 	 &0.370 		\\[0.5ex]
 \hline\\[-2ex]
 & & parallel & & \\
 \hline\\[-2ex]
 m9 &(4,0,0,3)	&0.542       &1.769 	 &0.539 		\\
 m10 &(9,0,0,3)	&1.216       &2.649 	 &0.592 		\\
 m11 &(11,0,0,3)	&1.622       &3.097 	 &0.664 		\\[0.5ex]
\hline
\hline
\end{tabular}
\vspace*{-0.4cm}
\begin{center}
\begin{minipage}{14cm}
\hspace*{3cm}
\caption{\small{Values for the RI$'$ scale defined as $a \bar\mu_0{=}\frac{2\pi}{L} (\frac{n_t}{2} {+} \frac{\pi}{4}, n_x,n_y,n_z)$,
where $L$ is the spatial extent of the lattice. The values are given in lattice and physical units. 
The last column corresponds to the ratio $\hat{P}$ defined in Eq.~(\ref{Phat}).}}
\label{tab1}
\end{minipage}
\end{center}
\end{center}
\vspace*{-0.5cm}
\end{table}
}}

\begin{figure}[!h]
\centering
\includegraphics[scale=0.285,angle=-90]{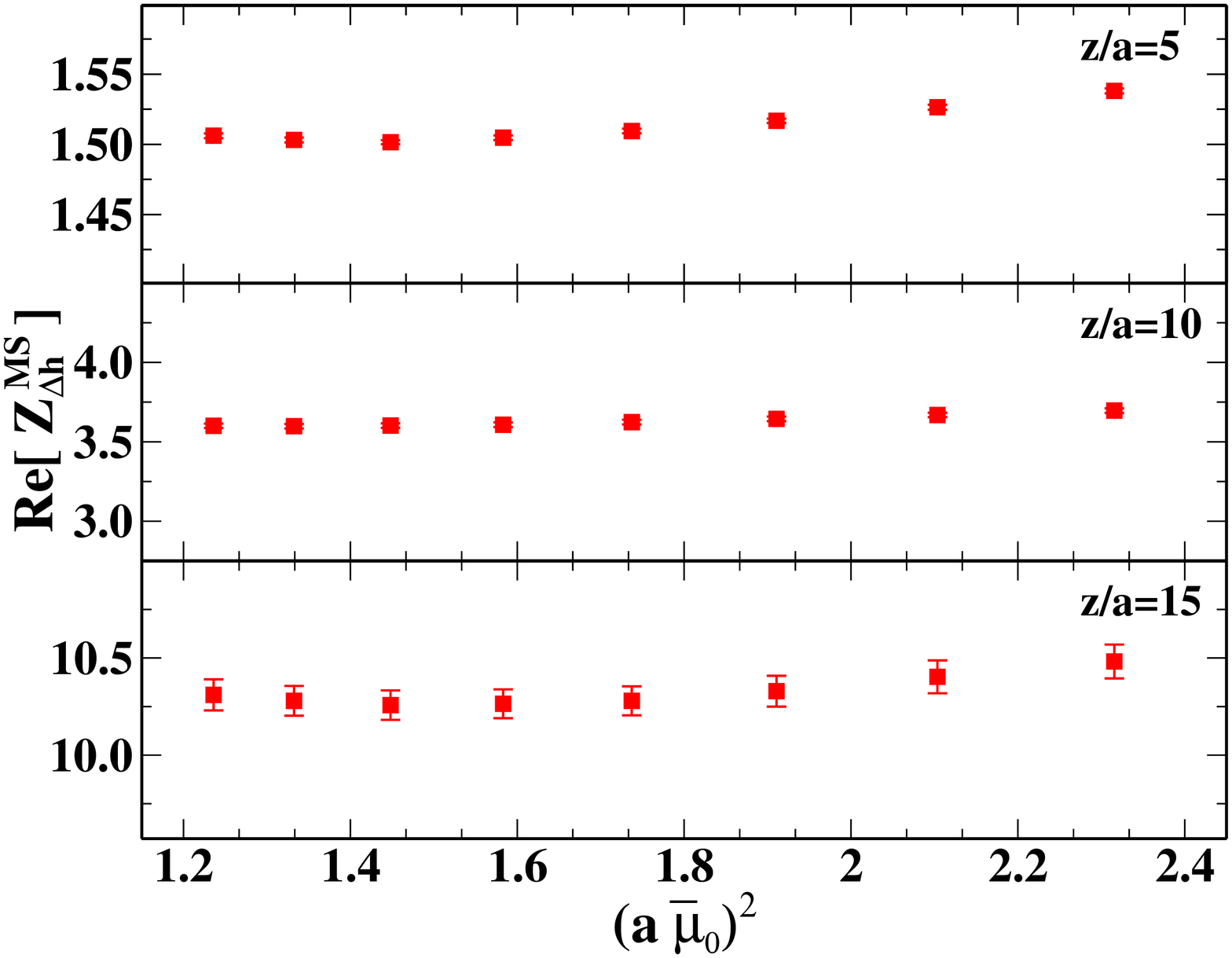}\,\,\,
\includegraphics[scale=0.285,angle=-90]{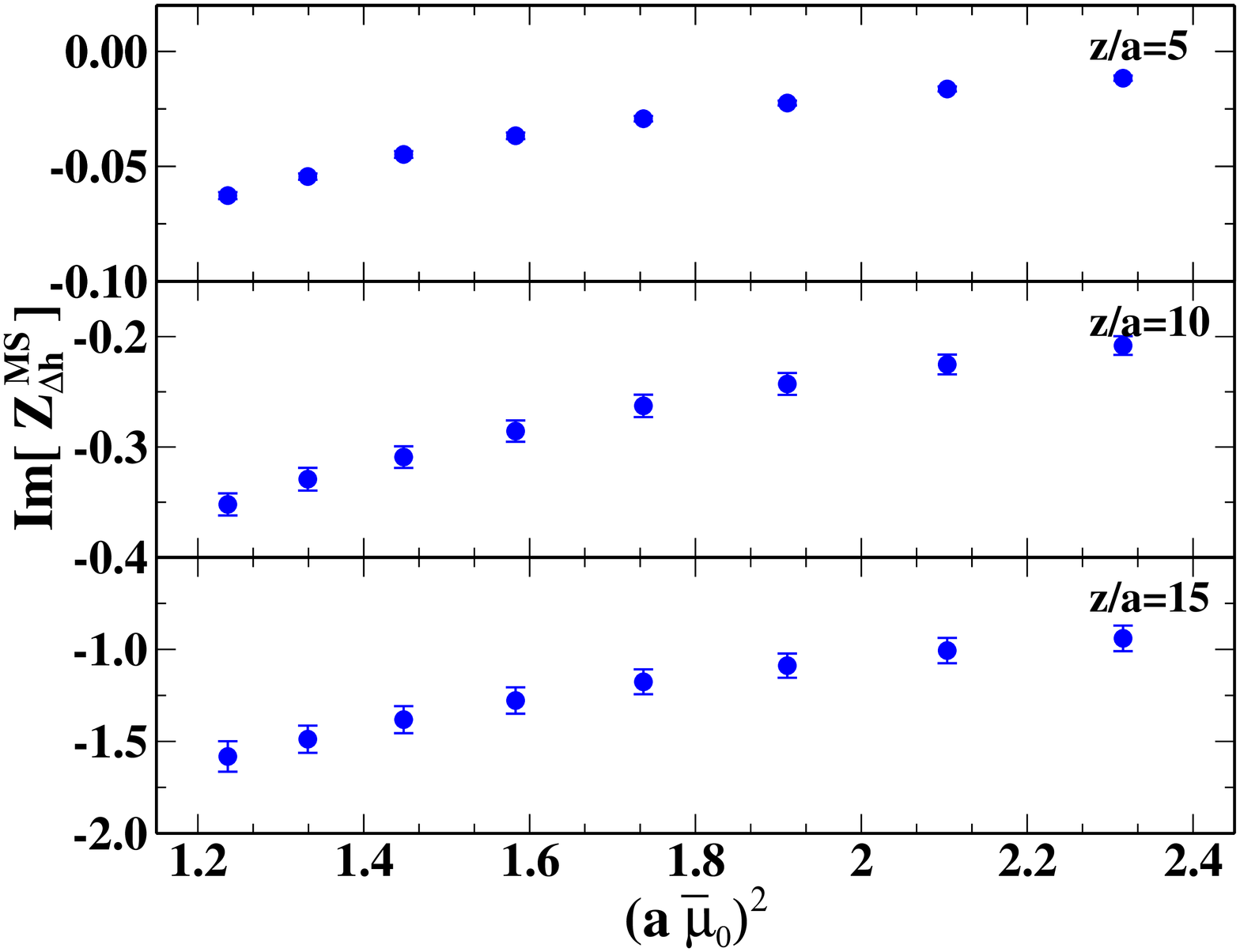}
\vspace*{-0.3cm}
\begin{minipage}{15cm}
\hspace*{3cm}
\caption{\small{Left: Real part of $Z^\MSb_{\Delta h}$ for scales labeled by m1$-$m8 and $z/a{=}5,10,15$.  
Right: Same as left panel for the imaginary part of $Z^\MSb_{\Delta h}$.}}
\label{fig:Za_vs_scale} 
\end{minipage}
\end{figure}

Since we present the renormalized matrix elements for the helicity PDF, we focus on $Z_{\Delta h}$ for this analysis. In Fig.~\ref{fig:Za_vs_scale},
we show $Z^{\MSb}_{\Delta h}$ for selected values of $z$ as a function of the RI$'$ scale, $(a\,\bar\mu_0)^2$, using the
diagonal values labeled by m1$-$m8. The real (imaginary) part is shown in the left (right) panel. We find a residual dependence on $(a\,\bar\mu_0)^2$, 
mostly affecting the imaginary part. This is due to the fact that the vertex function depends not only on the magnitude and direction of the renormalization
scale, especially the $z$-direction, as this is parallel to the Wilson line. Upon evolving to the same scale, the $Z$-factors should not depend on the initial scale
if the evolution is known to higher loops in perturbation theory and discretization effects are sufficiently small. The nonzero slope of the plots shown in 
Fig.~\ref{fig:Za_vs_scale} is an indication of non-negligible lattice artifacts, as well as, a consequence of the truncation of the conversion factor to one-loop level. 
However, further investigation is required in order to attempt disentangling the two effects. We address this in Subsection~\ref{sub3.2}.

\begin{figure}[!h]
\centering
\includegraphics[scale=0.275,angle=-90]{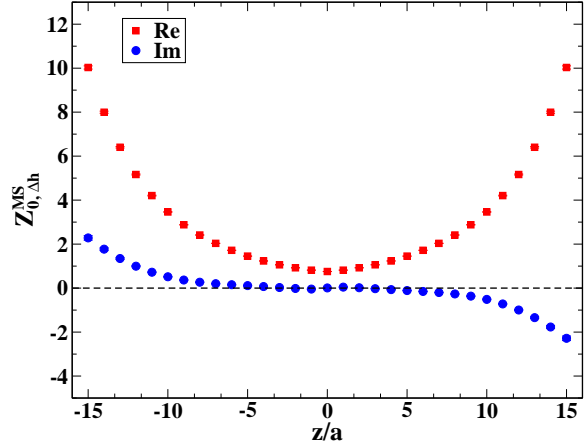}
\vspace*{-0.3cm}
\begin{minipage}{15cm}
\hspace*{3cm}
\caption{\small{Extrapolated $Z$-factor for the helicity operator, $Z_{0, \,{\Delta h}}^{\MSb}$, for 5 steps of HYP smearing. 
The employed fit range for the RI$'$ scale is $(a\,\bar\mu_0)^2\,\epsilon\, [1.4 {-} 2.0]$.}}
\label{fig:ZA_HYP5_extrapolated} 
\end{minipage}
\end{figure}

\vspace*{0.15cm}
\noindent
To remove the residual dependence on $(a\,\bar\mu_0)^2$ we fit $Z^{\MSb}_{\Delta h} $ with the function
\be
\label{eq:extrapol}
Z^{\MSb}_{\Delta h}  = Z_{0, \,{\Delta h}}^{\MSb}  + Z_{1, \,{\Delta h}}^{\MSb} \, .\,(a\,\bar\mu_0)^2
\ee
and we take $Z_{0, \,{\Delta h}}^{\MSb}$ as our final result. This process is done for each value of 
the length of the Wilson line $z$, and repeated for several fit ranges for the diagonal choices m1$-$m8 given in Table~\ref{tab1}.
For the extrapolated value $Z_{0, \,{\Delta h}}^{\MSb}$ at 2 GeV, we choose the one obtained in the range 
$(a\,\bar\mu_0)^2\,\epsilon\, [1.4,\,2.0]$. By excluding $(a\,\bar\mu_0)^2$ close to 1, we avoid contamination from non-perturbative 
effects. The choice for the above fit range also excludes the two higher scales that have $\hat{P}{>}0.3$, which may carry sizable lattice artifacts. 
As an additional check of the quality of the fit, we find that the $\chi^2/d.o.f$ is small for the chosen range.
In Fig.~\ref{fig:ZA_HYP5_extrapolated}, we plot the extrapolated value $Z_{0, \,{\Delta h}}^{\MSb}$ (for 5 steps of HYP smearing), which 
is applied on the bare nucleon matrix element of the helicity quasi-PDF in Subsection~\ref{sub3.2}.

\vspace*{0.5cm}

\begin{figure}[!h]
\centering
\includegraphics[scale=0.285,angle=-90]{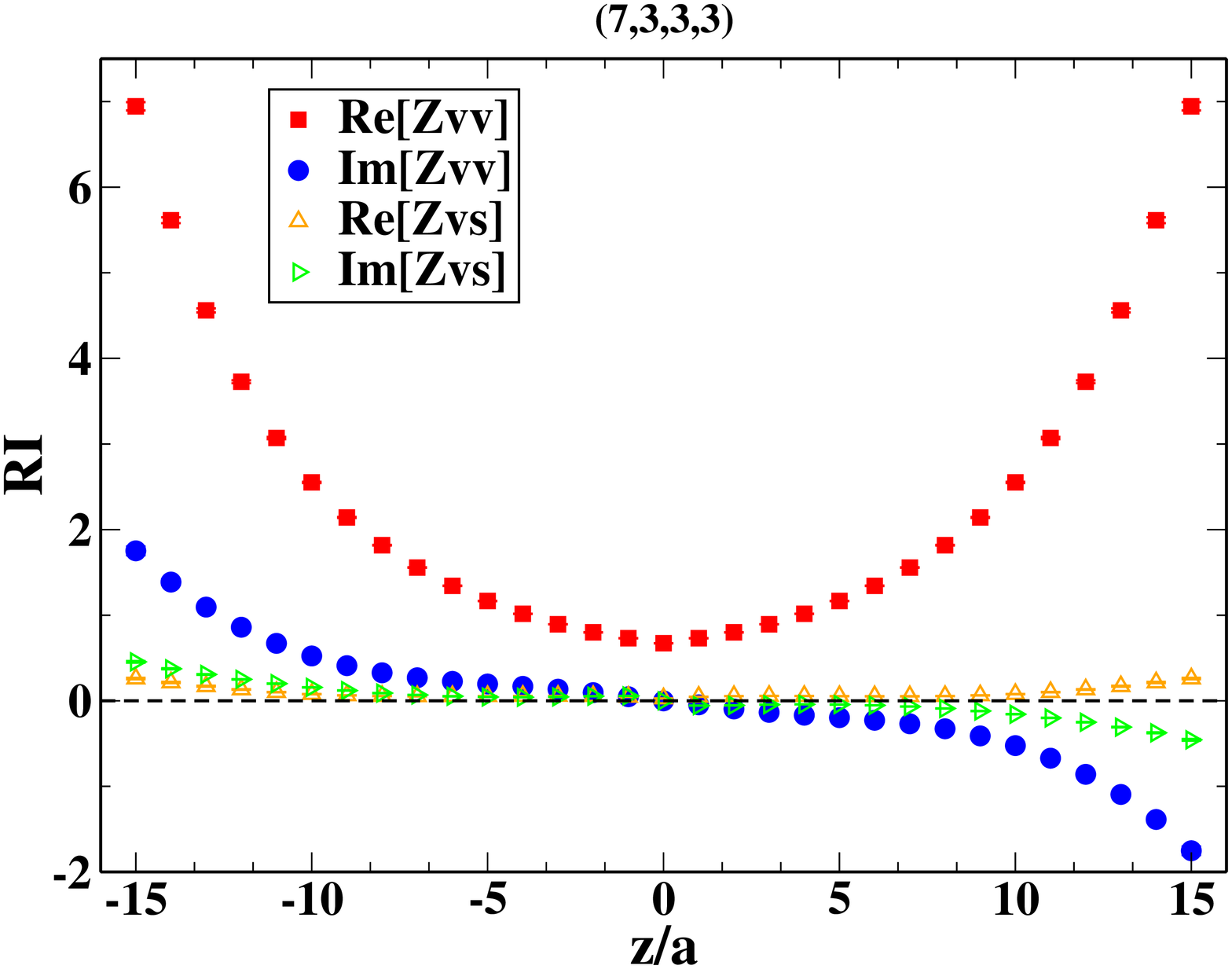}\,\,\,
\includegraphics[scale=0.285,angle=-90]{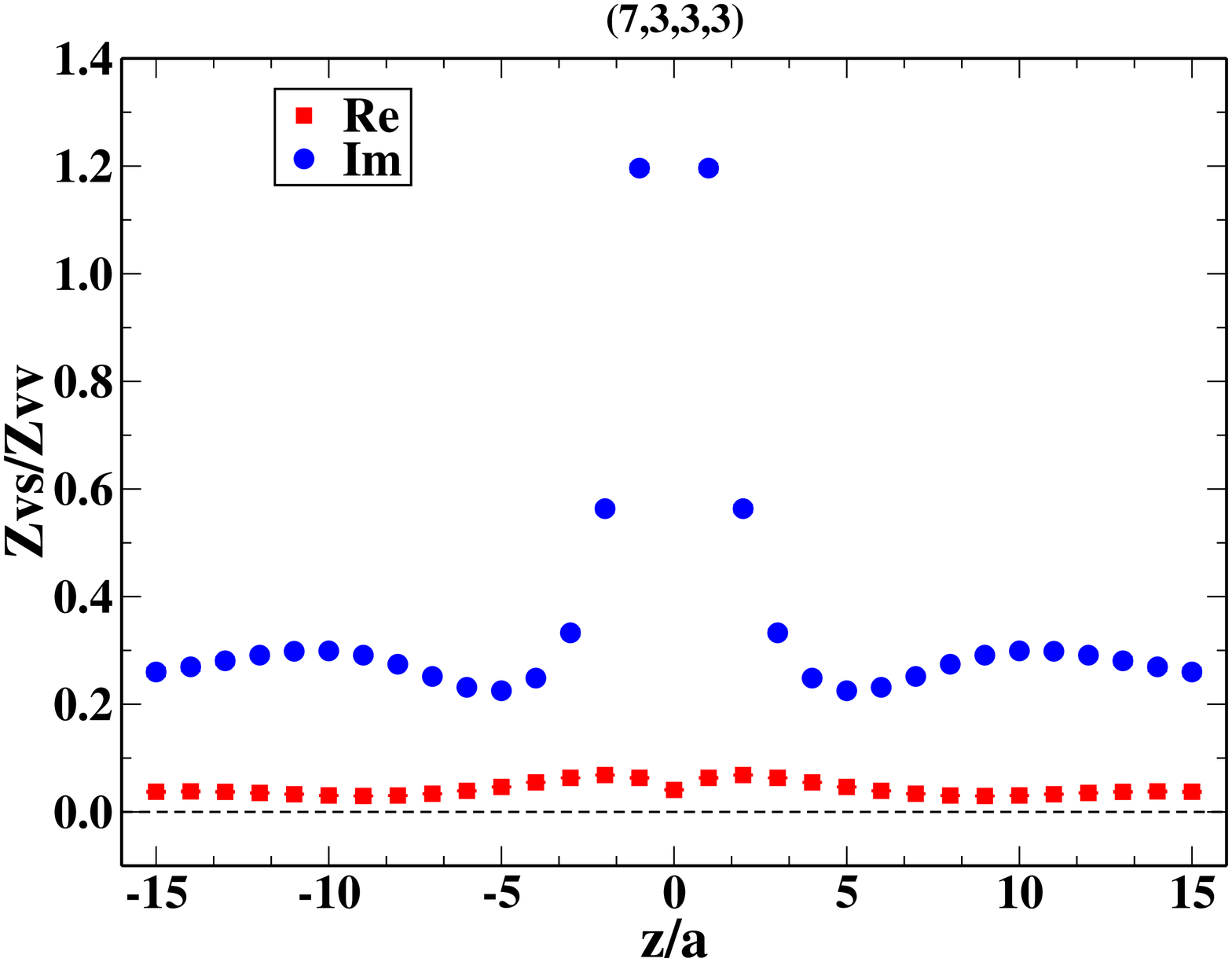}
\vspace*{-0.3cm}
\begin{minipage}{15cm}
\hspace*{3cm}
\caption{\small{Left: multiplicative ($Z^{\rm RI'}_{VV}$) and mixing  ($Z^{\rm RI'}_{VS}$) coefficients entering
the renormalization of the unpolarized quasi-PDF, in the RI$'$ scheme for the ``(7,3,3,3)'' scale. Right: 
ratio of the real (red squares) and imaginary (blue circles) parts of $Z^{\rm RI'}_{VS}$ over $Z^{\rm RI'}_{VV}$.}}
\label{fig:Zvv_Zvs} 
\end{minipage}
\end{figure}

\vspace*{0.5cm}
The prescription for obtaining $Z_{VV}$ and $Z_{VS}$ has also been applied on the same ensemble. 
We find that both the HYP smearing and the choice of ``diagonal'' scales suppresses the mixing
coefficients $Z_{VS}$ and $Z_{SV}$. In  the left panel of Fig.~\ref{fig:Zvv_Zvs}, we plot 
$Z^{\rm RI'}_{VV}$ and $Z^{\rm RI'}_{VS}$ for the scale  ``(7,3,3,3)'' and 5 steps of HYP smearing. 
The real and imaginary parts of $Z^{\rm RI'}_{VS}$ are of the same magnitude, but at 
least an order of magnitude smaller than the multiplicative factor $Z^{\rm RI'}_{VV}$. This can be
seen from the right panel of Fig.~\ref{fig:Zvv_Zvs}, where we plot the ratios 
$\Re[Z^{\rm RI'}_{VS}]/\Re[Z^{\rm RI'}_{VV}]$ (red squares) and 
$\Im[Z^{\rm RI'}_{VS}]/\Im[Z^{\rm RI'}_{VV}]$ (blue circles). From the left plot one observes that
the imaginary part of both $Z^{\rm RI'}_{VV}$ and $Z^{\rm RI'}_{VS}$ is compatible with zero for $|z/a|<4$.
Therefore, the large values of the blue points in the region $|z/a|<4$ in the right plot are no indication of significant 
mixing.

\subsection{Assessment of systematic uncertainties}
\label{sub3.2}
\medskip

In the framework of this study we have performed several investigations on the systematic uncertainties 
related to the truncation of the one-loop conversion factor, as well as, discretization effects. In this subsection
we present the main conclusions of this study, and we give estimates of these effects.

\begin{figure}[h!]
\centering
\includegraphics[scale=0.3,angle=-90]{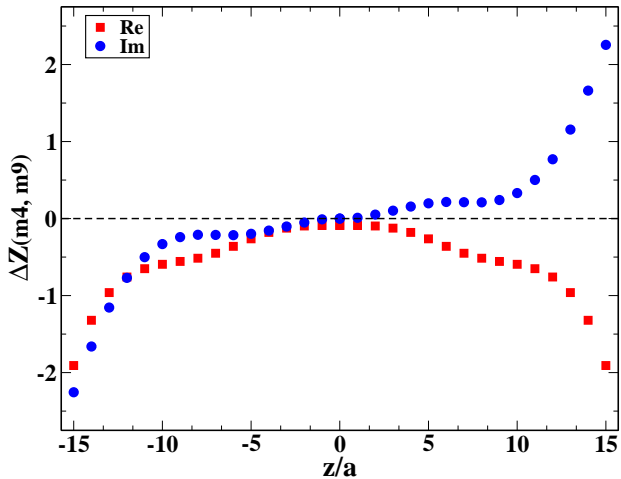}
\vspace*{-0.3cm}
\begin{minipage}{15cm}
\hspace*{3cm}
\caption{\small{Difference of $Z_{\Delta h}^\MSb$ between the ``parallel'' (m4) and ``diagonal'' (m9) cases for $\bar\mu_0$. 
The real and imaginary parts are shown with red squares and blue circles, respectively.}}
\label{fig:ZAdiff}
\end{minipage}
\end{figure}

\vspace*{0.3cm}
Prior converting to the $\MSb$ scheme, we compute the $Z$-factors for different values of the RI$'$ scale $\bar\mu_0$. 
For example, $Z_{\Delta h}^{\rm RI'}$ shown in the two plots of Fig.~\ref{fig:ZA} on the scales m4 and m9 is different. 
Upon evolving to the same scale, the extracted values should agree if the evolution has converged; typically this happens 
at two or three loops in perturbation theory. Thus, one should be able to compare the data extracted for 
$Z_{\Delta h}^\MSb$ at 2 GeV for the two scales presented in Fig.~\ref{fig:ZA}. The difference between the two, $\Delta Z({\rm m4, m9})$, 
is an indication of the presence of lattice artifacts (mainly in the ``parallel'' case), coupled with the truncation of the conversion factor. 
This is demonstrated in Fig.~\ref{fig:ZAdiff} for both the real and imaginary parts, and it is interesting to see that the difference 
increases as $z$ becomes larger. Based on this observation, we expect that such an increase of the lattice artifacts is also present 
for the ``diagonal'' case, but less severe. Another evidence of the presence of non-negligible systematics is the fact that the imaginary 
part of $Z_{\Delta h}^\MSb$ is nonzero. In Fig.~\ref{fig:ZA}, we find a small imaginary part for the ``diagonal'' scale, which has an 
increasing trend at large values of $z$.

\vspace*{0.3cm}

\vspace*{0.5cm}
Based on the above, we conclude that the systematic uncertainties related to lattice artifacts and the conversion factor must be addressed, 
in order to extract reliable estimates on the renormalization functions. It is our intention to reduce both effects in the near future, which 
will eliminate systematic uncertainties propagated to the estimates of the quasi-PDFs. 
Understanding the uncertainties dominating the large-$z$ region is crucial, as the matrix element for these values also enters in the Fourier 
transform that yields the quasi-PDF. Preliminary explorations indicate that a likely magnitude of the two-loop contribution might suppress 
the imaginary part of $Z_{\Delta h}^{\overline{\rm MS}}$. Even though we are currently in no position to accurately quantify these systematics, 
we will estimate upper bounds. In particular, we try to estimate the effect of each one individually.

\vskip 0.75cm
\centerline{\bf\underline{\textit{Lattice artifacts}}}
\vskip 0.25cm

The $Z$-factor for the helicity quasi-PDF at $z{=}0$ reduces to the renormalization function of the local axial current $Z_A$. 
$Z_A$ is scheme- and scale-independent, and thus, the case $z{=}0$ allows us to study the lattice artifacts.  $Z_A$ has been
already evaluated for this ensemble in Ref.~\cite{Alexandrou:2015sea}. In the latter calculation several diagonal scales have 
been employed and, together with a technique for the removal of the lattice artifacts, the extracted value was found to be 
$Z_A{=}0.7556(5)$. In the present calculation at $z{=}0$ we find a value of $Z_A{=}0.8620(15)$ using the scale m9 and 
$Z_A{=}0.7727(2)$ for the scale m4. This is yet another indication that lattice artifacts are large for the  ``parallel'' 
renormalization scale and less severe, but not negligible, for the ``diagonal'' case. 

\begin{figure}[h]
\centering
\includegraphics[scale=0.285,angle=-90]{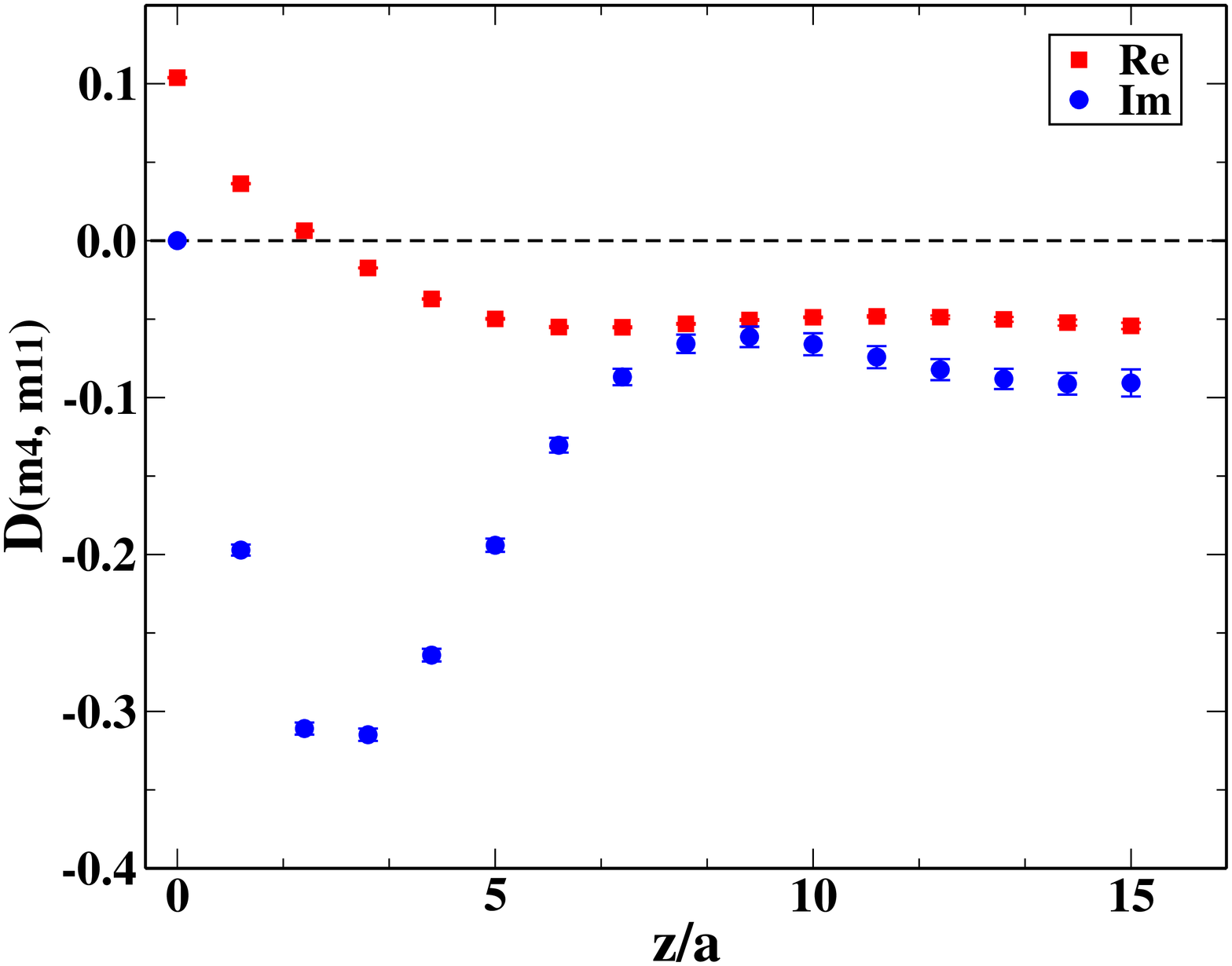}
\includegraphics[scale=0.285,angle=-90]{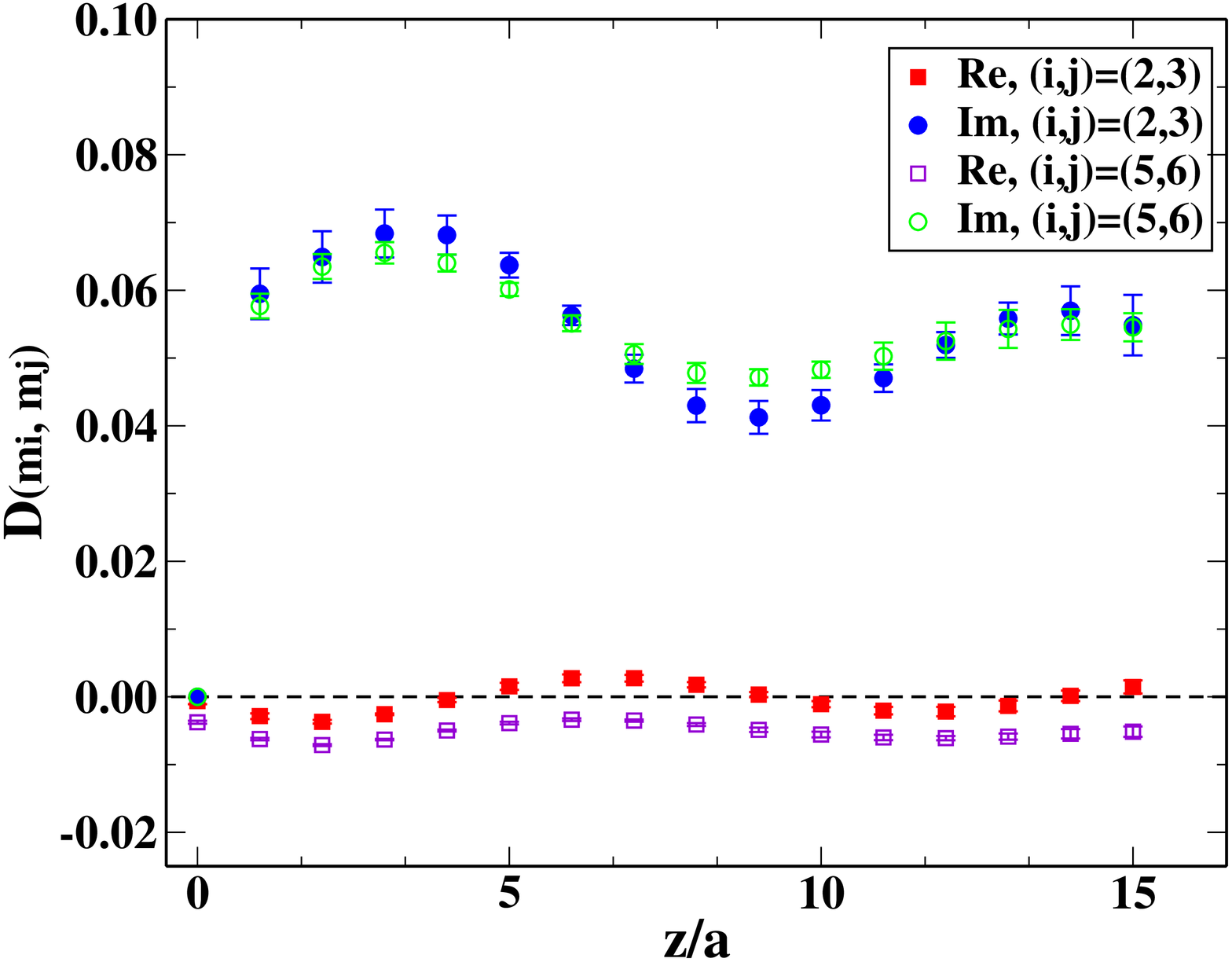}
\vspace*{-0.3cm}
\begin{minipage}{15cm}
\hspace*{3cm}
\caption{\small{Left: The ratio of Eq.~(\ref{eq:D}) using $(a\,\bar\mu_0)^2{=}1.583$ and $(a\,\bar\mu_0')^2{=}1.622$. 
Right: Similar to the left plot, for the pairs $( (a\,\bar\mu_0)^2, (a\,\bar\mu_0')^2 ){=}(1.332, 1.448),\, (1.737,1.911)$}}
\label{fig:ZA_diff_m4_m11} 
\end{minipage}
\end{figure}

An additional test that may be performed for any value of $z$ is the comparison of $Z_{\Delta h}^{\rm \MSb}$ between scales
with different components, but same $(a\,\bar\mu_0)^2$. For this, we utilize the values labeled as m10 and m11, which can be compared 
to m1 and m4, respectively. Since these pairs are approximately at the same value of $(a\,\bar\mu_0)^2$, we can compare them directly
in the RI$'$ scheme without any evolution\footnote{We confirmed numerically that the conversion factors from m1 to m10, and from m4 
to m11 deviates from unity by less than 1$\text{\textperthousand}$, so we ignore it.}, and the difference can be interpreted as lattice artifacts.
We observe similar
behavior for the two pairs and as a demonstration we plot in Fig.~\ref{fig:ZA_diff_m4_m11} the following ratio for each value of $z$
\be
\label{eq:D}
D_{\rm Re\,(Im)}(\bar\mu_0,\bar\mu_0')\equiv \frac{Z_{\rm Re\,(Im)}^{\rm RI'}(\bar\mu_0) - Z_{\rm Re\,(Im)}^{\rm RI'}(\bar\mu_0')}{Z_{\rm Re\,(Im)}^{\rm RI'}(\bar\mu_0)}\,.
\ee

\vspace*{0.25cm}
In the left panel we show the case where $\bar\mu_0$ and $\bar\mu_0'$ correspond to the values m4 and m11, respectively. 
The difference from zero can be attributed to lattice artifacts. We find that the $Z$-factor extracted from m11 deviates from the value extracted 
from m4 by 5-10\% in the real part and up to 30\% in the imaginary part. This deviation seems to stabilize after $z/a{\sim}8$ to 5\% (10\%) 
for the real (imaginary) part.

Note that the left plot of Fig.~\ref{fig:ZA_diff_m4_m11} gives the estimate of the discretization effects for the ``parallel'' 
case compared to the ``diagonal'' one. It would be interesting to compute $D_{\rm Re\,(Im)}$ between two neighboring ``diagonal'' 
momenta in order to understand the change in the artifacts. We choose 2 pairs (m2,m3) and (m5,m6) and we find that both 
$D_{\rm Re}(\rm{m2,m3})$ and $D_{\rm Re}(\rm{m5,m6})$ are less than 1\%, while $D_{\rm Im}(\rm{m2,m3})$ and $D_{\rm Im}(\rm{m5,m6})$
are of the order of 6\%. This is a confirmation that excluding ``parallel'' momenta in the fit of Eq.~(\ref{eq:extrapol}) reduces
significantly the discretization effects.

\newpage
\centerline{\bf\underline{\textit{Truncation effects}}}
\vskip 0.25cm

\begin{figure}[h]
\centering
\includegraphics[scale=0.285,angle=-90]{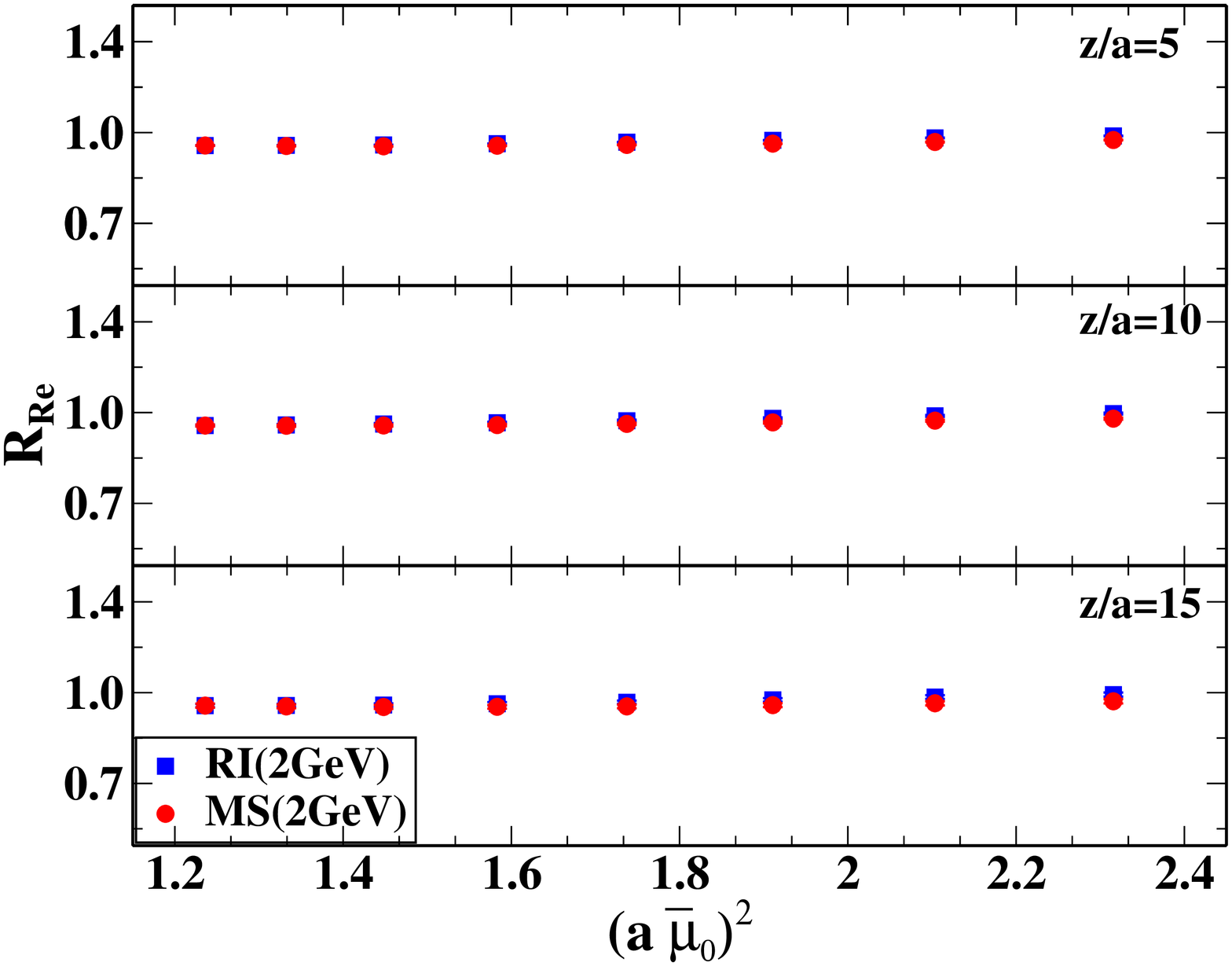}\,\,\,
\includegraphics[scale=0.285,angle=-90]{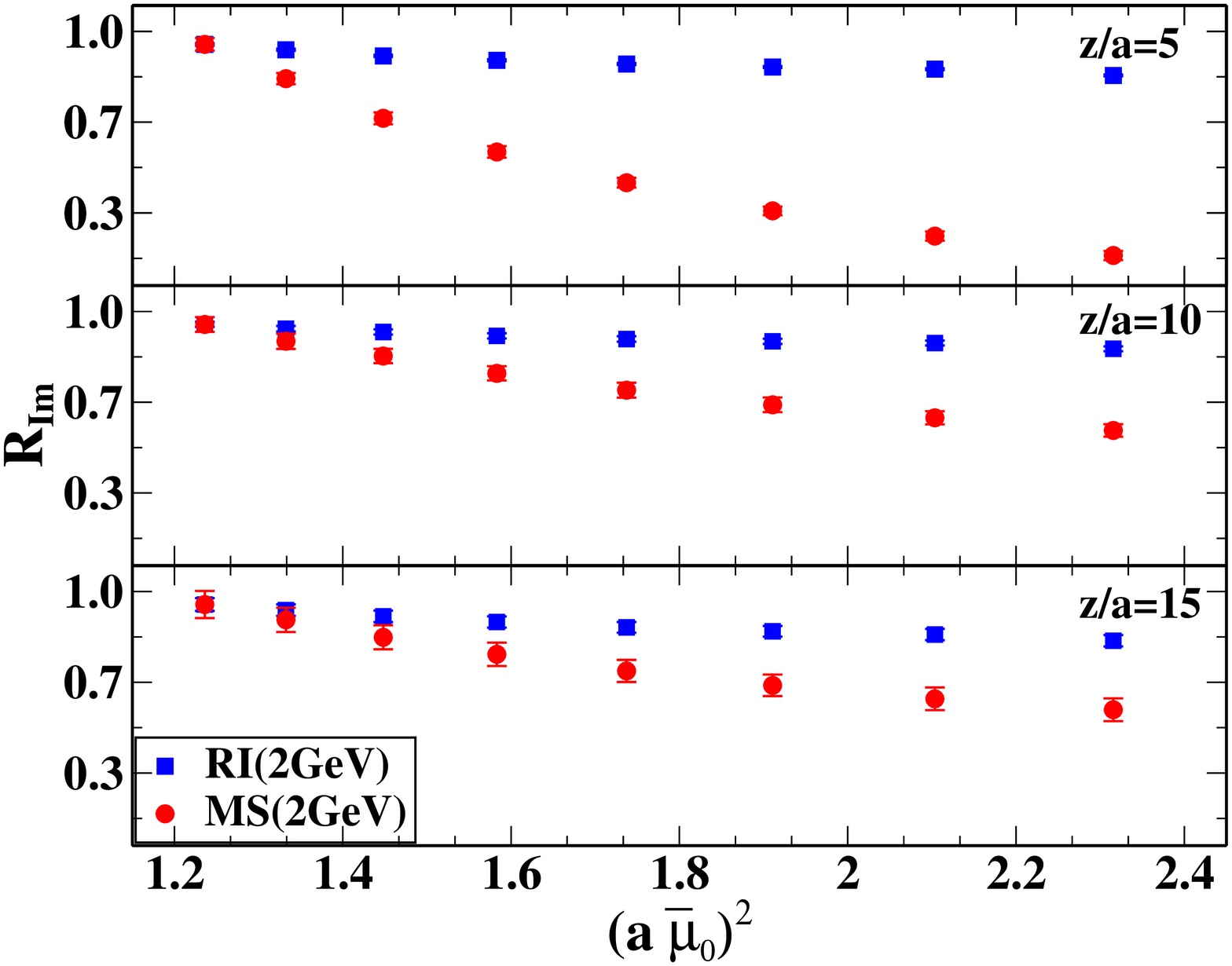}
\vspace*{-0.3cm}
\begin{minipage}{15cm}
\hspace*{3cm}
\caption{\small{The ratio of Eq.~(\ref{eq:R}) for $z/a{=}5,10,15$ as a function of the initial RI$'$ scale. 
All results have been evolved to 2 GeV and are normalized with the value of the $Z$-factor at 
$(a\bar\mu_0')^2{=}1.236$ (m1). Left: real part. Right: imaginary part. }}
\label{fig:RIvsMS} 
\end{minipage}
\end{figure}

In order to assess quantitatively the influence of the conversion factor that is known to one-loop perturbation theory, we form the ratios
\be
\label{eq:R}
R_{\rm Re\,(Im)}^{{\rm RI'}\,(\MSb)}(z, \bar\mu_0,\bar\mu_0';\bar\mu)\equiv 
\frac{Z_{\rm Re\,(Im)}^{{\rm RI'}\,(\MSb)}(z,\bar\mu_0;\bar\mu)}{Z_{\rm Re\,(Im)}^{{\rm RI'}\,(\MSb)}(z,\bar\mu_0';\bar\mu)}
\ee
both for the real and imaginary parts of the helicity $Z$-factor.  In Fig.~\ref{fig:RIvsMS}, we plot Eq.~(\ref{eq:R}) for the RI' and 
$\MSb$ case. These have been extracted at different values of $(a\bar\mu_0)^2$ (m1$-$m8) and evolved perturbatively to the same 
scale of $2$ GeV, using the results of Ref.~\cite{Constantinou:2017sej}. The ratio is always taken with respect to the 
smallest ``diagonal'' scale, m1. $R_{\rm Re\,(Im)}^{\rm RI'}$ depend on the truncation effects in the one-loop evolution to 2 GeV, while 
$R_{\rm Re\,(Im)}^\MSb$ is affected by scheme conversion truncation effects. Without contamination from lattice artifacts and 
truncation effects, Eq.~(\ref{eq:R}) should equal 1 in both schemes and for all values of $(a\bar\mu_0)^2$ and $z/a$. This realization 
allows for investigation of the truncation effects.

\bigskip

One observes that the imaginary part is more sensitive to the change of the initial RI$'$ scale, which is given in the x-axis.
To demonstrate this, we keep the range of y-axis the same for all plots of Fig.~\ref{fig:RIvsMS}.
The difference between the ratios of the real part extracted from different RI$'$ scales $(\bar\mu_0)^2$ is consistently 
small and reaches at most approx.\ 5\% in the RI$'$ scheme and 3\% in the $\MSb$ scheme, regardless of the Wilson line 
length, $z$.  As we mentioned above, for $z{=}0$ we can compare to the highly reliable value found in Ref.~\cite{Alexandrou:2015sea}; 
our current value from the scale m1 is around 2\% higher. Hence, we can estimate the typical size of 
discretization effects to be of order 2-5\% in the real part of the $Z$-factors.

The differences in the $\MSb$ ratios are combinations of lattice artifacts and the conversion truncation.
We observe that the conversion to the $\MSb$ scheme decreases the differences in the $\MSb$ scheme to at most 3\%.
Hence, truncation alone in the real part seems not to have an effect larger than 2\%. Since we know that conversion 
truncation effects are very small for small values of $z$ (in particular, they are zero for $z{=}0$, where $Z_A$ 
is scheme-independent), we can take 0-2\% as our estimate. In the end, given the fact that truncation 
effects seem to be opposite to the influence of lattice artifacts, we estimate that the total effects present in the 
real part of $Z_{\Delta h}^\MSb$ should not exceed 5\%.

The situation is somewhat different in the ratio of the imaginary parts shown in the right panel of Fig.~\ref{fig:RIvsMS}. 
We conclude that the uncertainty from lattice artifacts can be of the order of 10\% in the RI$'$ scheme for intermediate 
and large $z$ ($|z|/a\geq5$\,\footnote{For smaller values of $z$, the imaginary part is small in absolute terms and 
Eq.~(\ref{eq:R})  becomes meaningless, i.e.\ even small absolute changes of the imaginary part of $Z_{\Delta h}^\MSb$ 
can imply large changes in $R$.}). From the right panel of Fig.~\ref{fig:RIvsMS}, the total uncertainty in 
$R_{\rm Im}^\MSb$ can get enhanced to around 40\% for large and up to 80\% for intermediate Wilson line lengths. 
In addition, the total uncertainty in the imaginary part of $Z_{\Delta h}^\MSb$ is basically of the order of its magnitude,
as it is expected to be real. Note that ignoring the imaginary part does not provide any solution, as  
 they need to come out zero from the computation in order to claim that the computation is fully reliable.
Thus, conversion truncation effects become the most major source of uncertainty of the imaginary parts of the $Z$-factors.

\bigskip
Based on the study presented in this subsection, we present in Table~\ref{tab2} a summary of our quantitative estimates on 
the systematic uncertainties present in $Z_{\Delta h}^\MSb$. The estimate of the total effect is not a simple sum of 
isolated effects, but takes into account their different signs discussed above. For the imaginary part, the estimates are for 
$|z/a|\geq5$, since $\Im[Z_{\Delta h}^\MSb]{\sim}0$ for smaller values of $z$ and the relative effects become meaningless.

We would like to stress that due to the complex multiplication of the $Z$-factors and the bare matrix elements,
the large uncertainty in the imaginary part of the $Z$-factor implies that also real part of the renormalized matrix 
elements is affected. Furthermore, the uncertainties that we consider in the $Z$-factors translate differently to the 
uncertainties of the renormalized matrix elements for different Wilson line lengths and thus, we postpone the discussion of this 
influence to the next subsection.

{\small{
\begin{table}[!h]
\vspace*{0.25cm}
\begin{center}
\begin{tabular}{ccc}
\hline
\hline\\[-2ex]
Effect & $\Re[Z_{\Delta h}^\MSb]$   &  $\Im[Z_{\Delta h}^\MSb]$ \\[0.5ex]
\hline
\hline\\[-2ex]
Lattice artifacts & 2-5\% & $\lesssim10$\% \\[1ex]
Evolution truncation& 1-2\% & 1-2\% \\[1ex]
Conversion truncation& $\lesssim2$\% & $\lesssim100$\% \\[2ex]
\hline\\[-1.5ex]
Total & 3-5\% &  $\lesssim100$\% \\[1ex]
\hline
\hline
\end{tabular}
\vspace*{-0.4cm}
\begin{center}
\begin{minipage}{14cm}
\hspace*{3cm}
\caption{\small{Quantitative estimates of systematic uncertainties in the real and imaginary parts of $Z_{\Delta h}^\MSb$.}}
\label{tab2}
\end{minipage}
\end{center}
\end{center}
\end{table}
}}

\subsection{Renormalized results}
\label{sub3.2}

Once the $Z$-factors are obtained for the unpolarized, helicity and transversity quasi-PDFs, one may 
proceed with the application of the renormalization in the nucleon matrix elements. Here we mostly focus
on the helicity case, as the renormalization is multiplicative. The case of the unpolarized quasi-PDF is
briefly mentioned in the end of the subsection.

\begin{figure}[h]
\centering
\includegraphics[scale=0.285,angle=-90]{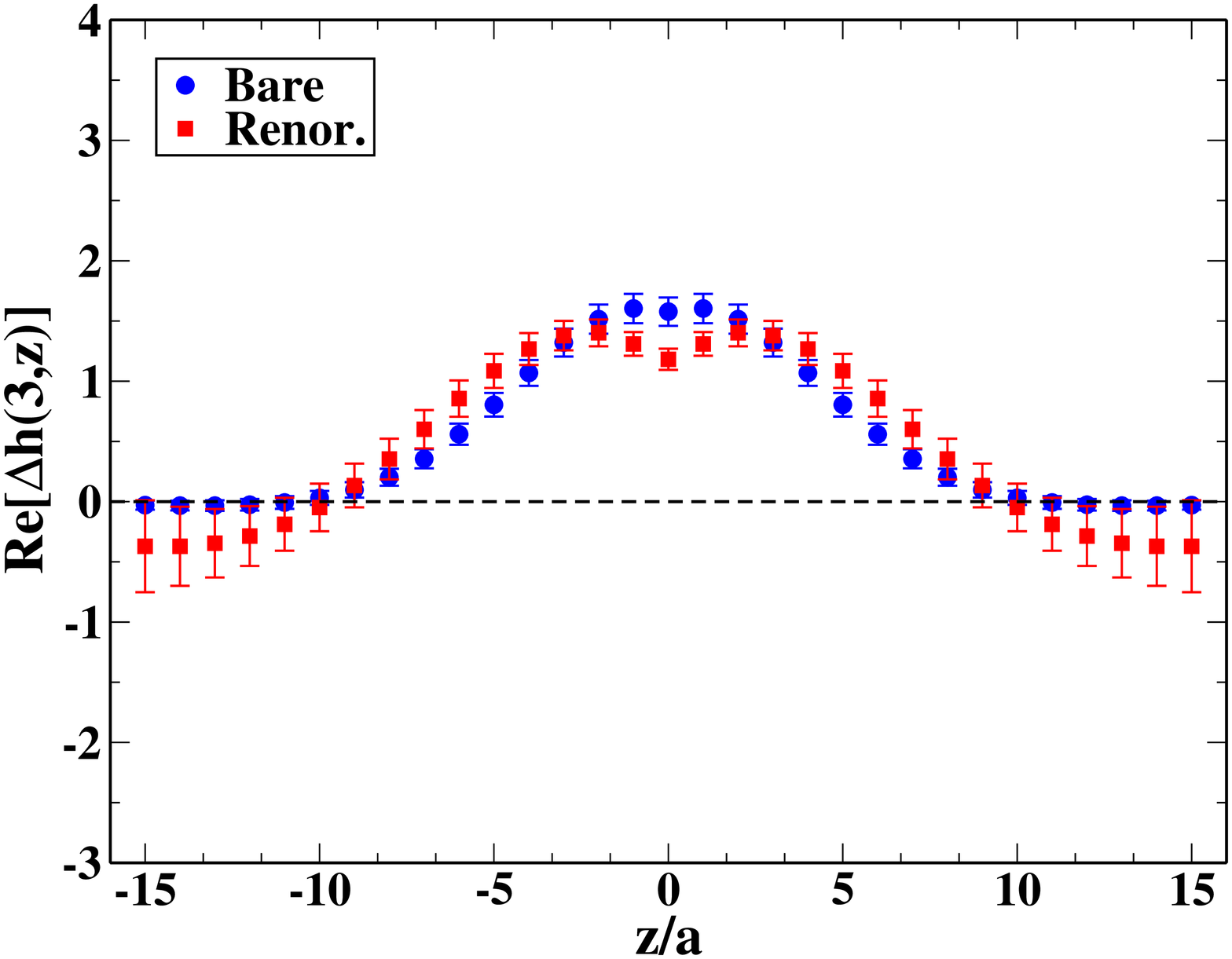}\,\,\,
\includegraphics[scale=0.285,angle=-90]{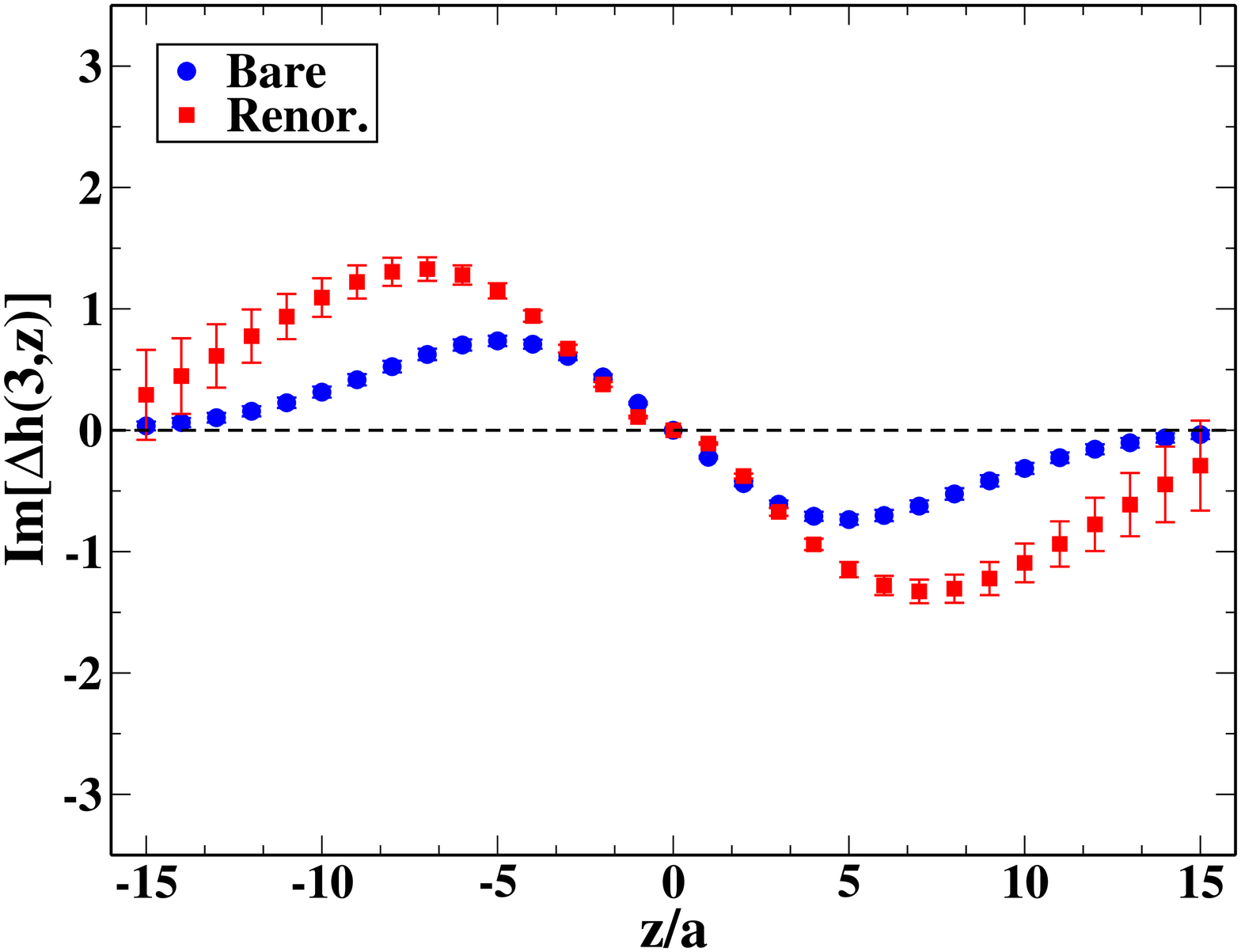}
\vspace*{-0.3cm}
\begin{minipage}{15cm}
\hspace*{3cm}
\caption{\small{Renormalized matrix elements for the helicity quasi-PDF in the 
$\MSb$ scheme at $\bar\mu{=}$2 GeV using $Z_{0, \Delta h}^{\MSb}$
extracted from the fit range $(a\,\bar\mu_0)^2\,\epsilon\, [1.4,\,2.0]$.}}
\label{fig:deltah} 
\end{minipage}
\end{figure}

In Fig.~\ref{fig:deltah}, we show the renormalized helicity nucleon matrix elements, and compare them with 
the bare ones. This is a straightforward procedure as there is no mixing for the axial operator and, therefore, 
the renormalization is only multiplicative:
\be
\Delta h^{\MSb} (z) = Z_{\Delta h}^\MSb(z)\cdot \Delta h^{bare}(z)\,\,.
\ee
The above formula involves complex quantities, and thus, $\Delta h^{\MSb} (z)$ is a mixture of the real and imaginary part
of the bare matrix element. However, each value of $z$ is renormalized independently.
We use $Z_{\Delta h}^\MSb$ from the extrapolation of Eq.~(\ref{eq:extrapol}) in the range 
$(a\,\bar\mu_0)^2\,\epsilon\, [1.4,\,2.0]$, as shown in Fig.~\ref{fig:ZA_HYP5_extrapolated}. 
One can see in Fig.~\ref{fig:deltah} that for small values of $z$ there is a slight suppression of 
the renormalized real part with respect to the bare one ($\Re [Z_{\Delta h}^\MSb<1]$). 
$\Re[\Delta h^{\MSb}]$ is compatible with zero for $|z/a|{>}8$, but with increased statistical 
uncertainties. The effect of the renormalization on the imaginary part of the matrix element is
profound in the large $z$ region, where we observe an amplification of its value and a shift of the
maximum to larger z, as compared to the bare one.
\bigskip

\vspace*{0.5cm}
The renormalized matrix elements of the helicity quasi-PDF are presented here for the first time, and they demonstrate the 
enormous progress in the field of the quasi-PDFs. However, before attempting to compare with the physical PDFs, we must 
understand the uncertainties that are inherited to $\Delta h^{\MSb} (z) $ from its renormalization function.
As we argued in the previous subsection, a robust computation needs the subtraction of $\mathcal{O}(g^2\,a^\infty)$ 
lattice artifacts and a significant reduction of truncation uncertainties in the conversion between the RI$'$ and $\MSb$ schemes.
We attempt setting bounds on these systematics, starting with the real part of renormalized matrix elements, which reads:
\be
\label{eq:Re}
\Re[\Delta h^\MSb]=\Re[Z_{\Delta h}^{\MSb}]\,\, \Re[\Delta h^{bare}]-\Im[Z_{\Delta h}^{\MSb}]\,\, \Im[\Delta h^{bare}]\,.
\ee
For small values of $z$, $\Im[Z_{\Delta h}^{\MSb}]$ is approximately zero and thus $\Re[\Delta h^\MSb]$ is approximately equal to the first term of Eq.~(\ref{eq:Re}).
On the contrary, for $|z/a|\gtrsim10$, $\Re[\Delta h^\MSb]$ receives significant contributions from the imaginary part, because $\Re[\Delta h^{bare}]<\Im[\Delta h^{bare}]$.

The uncertainty in the small-$z$ region of renormalized matrix elements is dominated by uncertainties in the real part of the $Z$-factor, which, as argued in the previous subsection, should not exceed 5\%. 
The local minimum observed at $z{=}0$ is likely to result from lattice artifacts and truncation effects.
In the large-$z$ region, the real part of the renormalized matrix elements receives negative contributions from the imaginary part of the bare matrix element if $Z_{\Delta h}^{\MSb}$ has a non-zero imaginary part and $\Im[\Delta h^{bare}]$ decays to zero more slowly than $\Re[\Delta h^{bare}]$, which is what we observe in the data.
This effect is unphysical and is expected to be strongly suppressed or eliminated by extending our calculation to the two-loop order in the conversion and by subtracting lattice artifacts.

The imaginary part of renormalized matrix elements,
\be
\label{eq:Im}
\Im[\Delta h^\MSb]=\Re[Z_{\Delta h}^{\MSb}]\,\, \Im[\Delta h^{bare}]+\Im[Z_{\Delta h}^{\MSb}]\,\, \Re[\Delta h^{bare}],
\ee
is enhanced in the intermediate-$z$ regime, compared to the bare matrix elements. This results mostly from the fact that 
$\Re[Z_{\Delta h}^{\MSb}]$ increases with increasing $z$ at a faster rate, than the decay of $\Im[\Delta h^{bare}]$.
However, the obtained values receive also contributions from the second term of Eq.~(\ref{eq:Im}), where $\Im[Z_{\Delta h}^{\MSb}]$ 
is subject to a large uncertainty, as discussed in the previous subsection.

We want to stress that it is not possible to give a single number for the relative uncertainty of the real and imaginary parts of $\Delta h^\MSb$.
According to Eqs.~(\ref{eq:Re}) - (\ref{eq:Im}), different regions of $z$ are influenced in a different way by the real and imaginary parts of the $Z$-factors.
For small $z$, where $\Im[Z_{\Delta h}^{\MSb}]$ is small, the propagated uncertainty in the matrix elements is dominated by the uncertainty of
 $\Re[Z_{\Delta h}^{\MSb}]$, which is of the order of 5\% and comparable to the currently attained statistical uncertainty. Thus, $\Delta h^\MSb$
is rather robust in this region. However, when $z$ is increased, the uncertainty from $\Im[Z_{\Delta h}^{\MSb}]$ starts to dominate and reaches 
100\% for large $z$. In this way, the improvements expected by the perturbative subtraction of $\mathcal{O}(g^2\,a^\infty)$ lattice artifacts and the extension 
to the two-loop perturbative conversion to the $\MSb$ scheme are very important for obtaining meaningful values of renormalized matrix elements and hence, also quasi-PDFs.

\vspace*{0.5cm}
The values of the multiplicative vector renormalization factor and the mixing coefficient can be used to properly 
renormalize the unpolarized quasi-PDF through Eq.~(\ref{h_R}). The successful renormalization requires the bare 
nucleon matrix elements for the scalar and vector operators, in order to extract the renormalized $h_V^{\rm RI'}$.
Then, it must be multiplied by the conversion factor $C_V$ to bring the results to the $\MSb$ scheme. 
Note that once the mixing is treated, the conversion factor is multiplicative and not a $2{\times}2$ matrix. 
This is due to the fact that no mixing is present in the continuum dimensional regularization.

\subsection{Matching to light-front PDFs}
Having the renormalized matrix elements, one can perform a Fourier transform and obtain the renormalized quasi-PDF, 
which represents the distribution of quark momenta for a finite-momentum nucleon moving in the chosen spatial ($z$) direction:
\begin{equation}
\label{qpdf}
\tilde{q}(x,\mu,P_3) = \int_{-\infty}^\infty \frac{dz}{4\pi} e^{-i z x P_3} \langle N |
\bar{\psi}(0,z)\Gamma W(z) \psi(0,0) |N\rangle^{\MSb,\mu}\,,
\end{equation}
where $x$ is the quark momentum fraction.
The quasi-PDF, expressed in our case in the $\MSb$ scheme at $\mu{=}2$ GeV, can then be connected to the light-front PDF, in the same scheme and at the same scale, using one-loop perturbative matching.
The matching procedure uses the fact that only the ultraviolet physics is different in quasi- ($\tilde{q}$) and light-front ($q$) PDFs \cite{Xiong:2013bka}.
Hence, the one-loop difference between them is expressed as the difference between vertex corrections (denoted $Z^{(1)}$ below) and wave function corrections ($\delta Z^{(1)}$) for the finite momentum and infinite momentum cases.
The generic matching formula for quasi-PDFs is: \cite{Ji:2015jwa,Chen:2016fxx}
\begin{eqnarray}
\label{invq}
q\left(x,\mu\right) &=& \tilde{q}(x,\mu,P_3) - \frac{\alpha_s(\mu)}{2\pi} \,\tilde{q}(x,\mu,P_3)\,\delta Z^{(1)}\!\left( \frac{\mu}{P_3}\right)  \nonumber\\*
& & - \frac{\alpha_s(\mu)}{2\pi} \int_{-\infty}^{\infty} Z^{(1)}\!\left( \xi, \frac{\mu}{P_3} \right) \tilde{q}\! \left( \frac{x}{\xi},\mu,P_3 \right) \frac{d\xi}{|\xi |} + \mathcal{O}(\alpha_s^2),
\end{eqnarray}
where $\alpha_s(\mu)$ is the strong coupling constant at the scale $\mu$.
In the integral, we exclude a small region around $\xi=0$, such that the argument of $\tilde{q}$ is not too large, as we would then pick up contributions from the mirror images of $\tilde{q}$ resulting from the periodicity of the Fourier transform in Eq.~(\ref{qpdf}).
The linearly divergent terms $\propto\Lambda/P_3$ ($\Lambda$: transverse momentum cutoff) of the matching 
formulae from Ref.~\cite{Xiong:2013bka} are not present in our renormalized results.

\bigskip
\begin{figure}[h]
\centering
\includegraphics[scale=0.55,angle=-90]{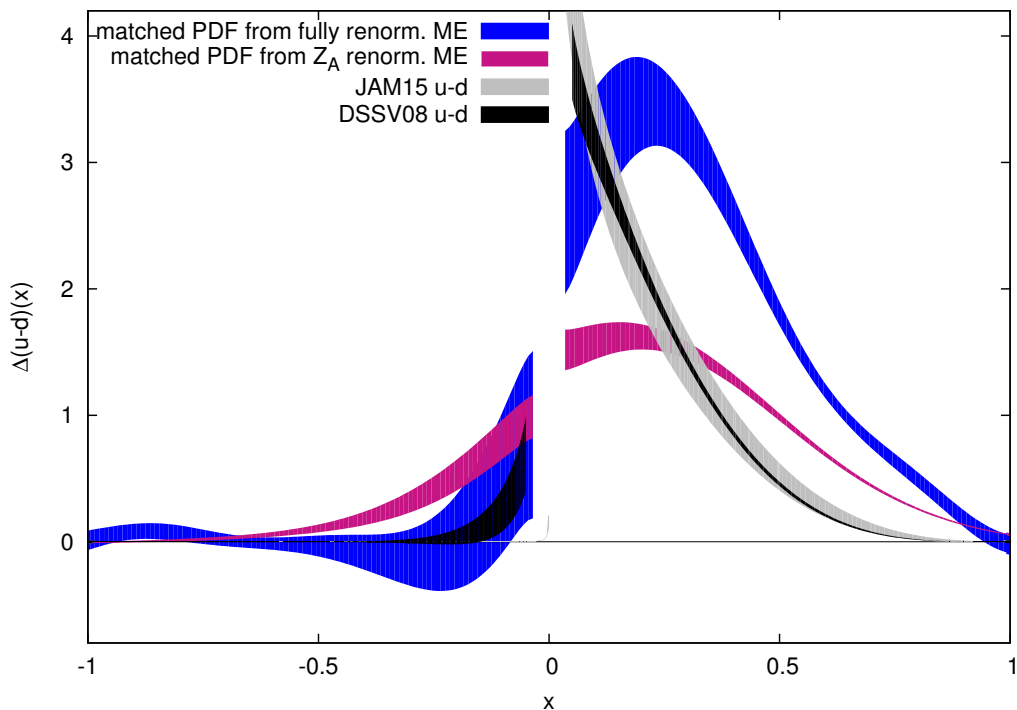}\,\,\,
\vspace*{-0.3cm}
\begin{minipage}{15cm}
\hspace*{3cm}
\caption{\small{Comparison of matched helicity PDF obtained from quasi-PDF computed with either fully renormalized matrix elements (blue) 
or with bare matrix elements multiplied by the local ($z{=}0$) axial current $Z$-factor, $Z_A$ (magenta). For purely orientational purposes, 
we also plot phenomenological PDFs (DSSV08 \cite{deFlorian:2009vb} and JAM15 \cite{Sato:2016tuz}).
However, we emphasize that no quantitative comparison with our results is aimed at, since careful consideration of a number of systematic effects is still needed. These include:
cut-off effects, non-physical pion mass, finite volume effects, possible contamination by excited states, extrapolation to infinite nucleon boost, as well as the improvements in the computation of $\MSb$ renormalization functions, postulated in the previous subsection.}}
\label{fig:matched} 
\end{minipage}
\end{figure}

\medskip
In Fig.~\ref{fig:matched}, we show the matched helicity PDF computed with either fully renormalized matrix elements obtained in this work (blue) or with bare matrix elements multiplied by the local ($z{=}0$) axial vector current renormalization function $Z_A$ (magenta), corresponding to our results from Ref.~\cite{Alexandrou:2016jqi}.
We observe that the renormalized matrix elements from this work move towards the phenomenological PDFs, which is promising.
In particular, the antiquark asymmetry is not overestimated any longer and actually this asymmetry becomes compatible with zero under current uncertainties.
In the quark part, there is an enhancement of the matched PDF for all values of $x$.
We emphasize again that the comparison with the phenomenological PDFs should be understood as only qualitative.
For quantitative comparison, a careful investigation of a number of systematic effects is still needed. These include:
cut-off effects, non-physical pion mass, finite volume effects, possible contamination by excited states, extrapolation 
to infinite nucleon boost, as well as the improvements in the computation of $\MSb$ renormalization functions, 
postulated in the previous subsection: subtraction of lattice artifacts computed in lattice perturbation theory and 
reduction of truncation effects in the perturbative conversion to the $\MSb$ scheme.

\section{Conclusions and discussion}

In this work we have presented a concrete prescription to renormalize non-perturbatively the matrix elements needed 
for the computation of quasi-PDFs. The employed scheme is RI$'$, which is then converted to the $\MSb$ scheme and evolved to 2 GeV; this is done
perturbatively to one-loop. We have argued that the renormalization condition properly handles both kinds of divergences present in the 
matrix elements: the standard logarithmic divergence and the power divergence specific to non-local operators containing a Wilson line. Furthermore,
we provide the renormalization conditions to eliminate the mixing in the case of the unpolarized quasi-PDF that mixes with the twist-3 scalar operator.

\medskip
We have also demonstrated the implementation of the proposed prescription to the helicity quasi-PDF and presented the corresponding renormalized 
matrix elements. This has allowed us to draw conclusions how to make the computation more robust, which is the main outcome of this work.
\begin{itemize}
\item First, an essential ingredient of a computation with controlled systematic uncertainties is the subtraction of one-loop lattice artifacts in the framework 
of lattice perturbation theory.
Following the ideas of Ref.~\cite{Alexandrou:2015sea}, we are currently computing the $\mathcal{O}(g^2\,a^\infty)$ artifacts that will be subtracted 
from the non-perturbative estimates for the $Z$-factors. In this way, the presence of large cut-off effects in the renormalized functions (especially
for ``parallel'' momenta) will be avoided to a large extent. 
\item Second, the conversion factor from the RI$'$ renormalization scheme to the $\MSb$ 
scheme is likely to have sizable higher order corrections that, among others, are responsible for the unphysical feature of the real part of the renormalized 
matrix element becoming negative for large Wilson line lengths. A two-loop computation of this conversion factor is expected to resolve this issue 
to a sufficient degree. We have performed numerical experiments that indicate that a natural change of the conversion factor by two-loop contributions, 
i.e.\ around 10-20\% (which is approximately $\alpha_s$ at the considered scale), should be enough to suppress the unwanted effect. A perturbative
calculation of the conversion factor to two loops is quite laborious and will be presented separately.
\end{itemize}

\medskip
To summarize, the renormalization program presented in this work together with future improvements that are being pursued,
will provide reliable estimates for the renormalization functions of the Wilson line fermion operators. In this fashion, the obtained 
renormalized matrix elements can be used as an input to 
calculate the quasi-PDFs and match them to light-front PDFs, which is the main aim of the whole approach.
Apart from the helicity case discussed in this work, we will address the transversity PDF in an analogous manner. For the unpolarized case, one needs
to take into account the mixing with the scalar operator, as explained and numerically demonstrated here. 
With this work, we have proposed and discussed a complete renormalization program of the quasi-PDFs, which has been a major 
uncertainty prior to this work and constitutes a crucial milestone in connecting lattice QCD results to the light-cone PDFs.

\vspace*{1cm}
\centerline{\bf\large{Acknowledgements}}
\medskip
We would like to thank the members of ETMC for useful and fruitful discussions. 
We also thank Rainer Sommer for discussions related to the arbitrary scale in the renormalization prescription. 
An award of computer time was provided by the INCITE program. This research also used resources of the Oak Ridge Leadership Computing Facility, which is a DOE Office of Science User Facility supported under Contract DE-AC05-00OR22725.
KC was supported in part by the Deutsche Forschungsgemeinschaft (DFG), project nr. CI 236/1-1.
MC acknowledges financial support by the U.S. Department of Energy, Office of Science, Office of Nuclear Physics, 
within the framework of the TMD Topical Collaboration, as well as, by the National Science Foundation under Grant No. PHY-1714407.
We acknowledge funding from the European Union's Horizon 2020 research and innovation program
under the Marie Sklodowska-Curie grant agreement No 642069. 
 
\vspace*{1cm}
\bibliographystyle{elsarticle-num}
\bibliography{references.bib}

\end{document}